\documentclass[11pt]{article}

\usepackage[preprint]{acl}

\usepackage{times}
\usepackage{latexsym}

\usepackage[T1]{fontenc}

\usepackage[utf8]{inputenc}

\usepackage{microtype}

\usepackage{inconsolata}

\usepackage{graphicx}
\usepackage{booktabs}
\usepackage{multirow}
\usepackage{adjustbox}
\usepackage[most]{tcolorbox}
\usepackage{colortbl}
\usepackage{enumitem}
\usepackage{tcolorbox}
\tcbuselibrary{breakable}
\usepackage{cuted}
\usepackage{hyperref}

\title{MetaCLASS: Metacognitive Coaching for Learning
with Adaptive Self-regulation Support}

\author{Naiming Liu \\
  Rice University \\
  \texttt{nl35@rice.edu} \\\And
  Richard Baraniuk \\
  Rice University \\
  \texttt{richb@rice.edu} \\\And
  Shashank Sonkar \\
  University of Central Florida\\
  \texttt{shashank.sonkar@ucf.edu} \\}

\begin{document}
\maketitle
\begin{abstract}
Large language models can generate fluent explanations, but effective tutoring requires supporting the learner's \emph{thought process}, not just delivering content. Metacognitive tutoring targets this gap by prompting \emph{planning}, \emph{monitoring}, \emph{debugging}, and \emph{evaluation}, and, crucially, deciding when to be active versus minimally present, based on learner signals and trajectory.
We introduce \textbf{MetaCLASS}, a learning-science grounded framework that formulates metacognitive tutoring as \emph{move selection} over 11 interpretable actions aligned to self-regulated learning (SRL) processes. MetaCLASS uses a two-phase procedure that first plans a pedagogical trajectory conditioned on learner profiles (calibration, help-seeking) and then generates natural dialogue consistent with that plan. This process yields 1,015 conversations spanning 7,711 turns, each annotated with turn-level metacognitive labels and supported by comprehensive validation analyses.
We benchmark nine open-source LLMs on predicting the next coach move given a question and dialogue context. The best model achieves only 43.2\% accuracy in coach move prediction, and models exhibit \emph{compulsive intervention bias}: in turns where effective metacognitive tutoring requires silence (41.7\% of cases), models predict ``no intervention'' only 6.6\% of the time, while severely over-predicting high-intervention moves. These results show that traditional content-based tutoring ability does not translate to metacognitive tutoring competence, positioning MetaCLASS as a testbed for developing tutors that promote SRL\footnote{Dataset and code is available \href{https://github.com/lucy66666/MetaCLASS}{here}.}.
\end{abstract}

\begin{figure*}[t]
\centering
\includegraphics[width=\linewidth]{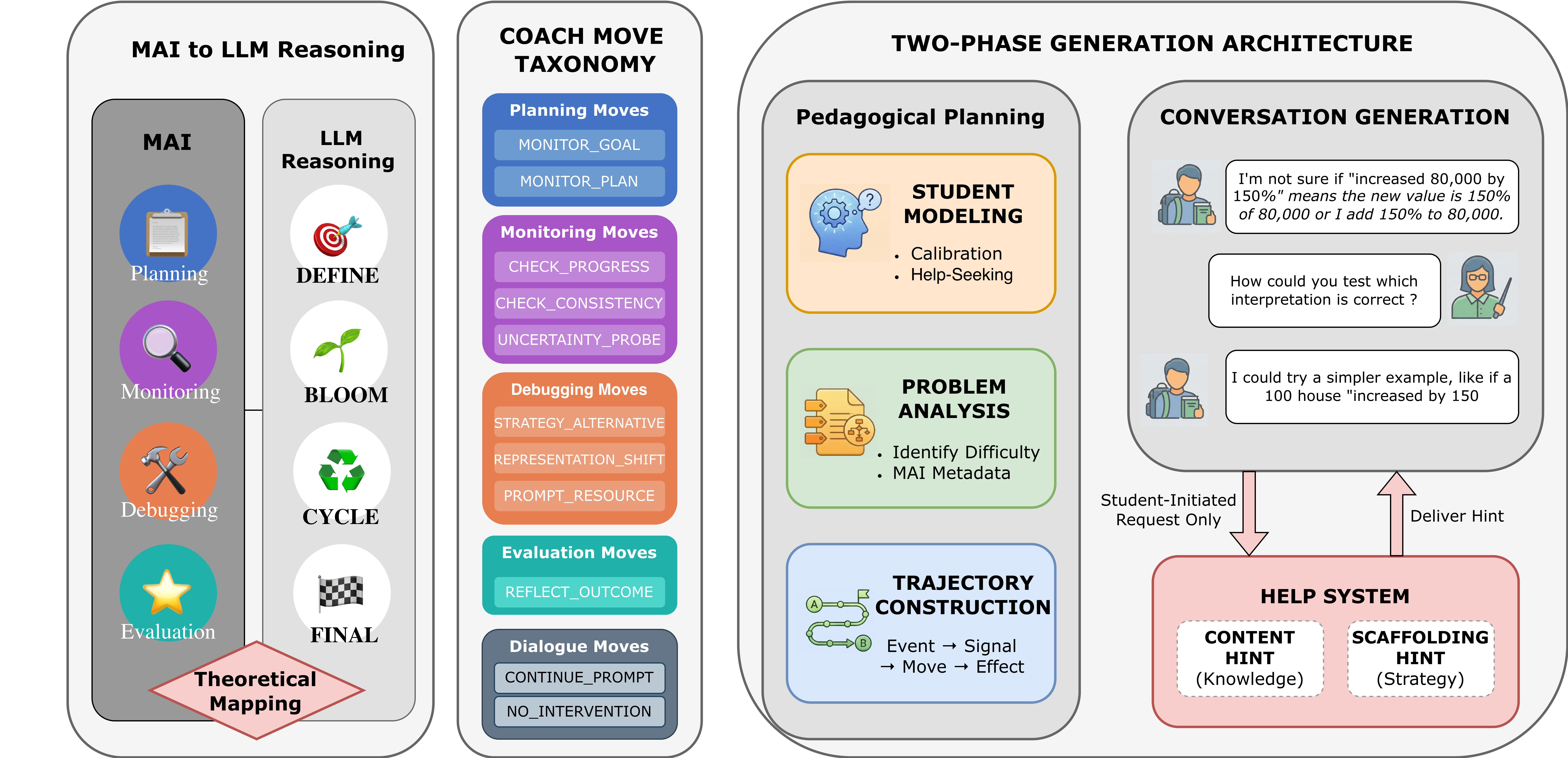}
\caption{An overview of the MetaCLASS Framework. \textbf{(Left)} We ground the framework in learning science by mapping MAI factors to LLM reasoning phases. \textbf{(Center)} We define a taxonomy of 11 interpretable coach moves, categorized by regulatory process. \textbf{(Right)} A two-phase generation architecture first plans a pedagogical trajectory together with learner profiles and problem analysis, which then guides the generation of metacognitive coaching dialogues.}
\label{fig:diagram}
\end{figure*}

\section{Introduction}

Modern AI tutoring systems can generate accurate explanations across domains, but often fail at a core tutoring skill: supporting the learner’s \emph{thought process}. The failure is not conceptual, but pedagogical. Good tutoring is not always the same as delivering correct content. It is the art of steering a learner's \emph{process}: clarifying goals, choosing strategies, monitoring understanding, recovering from confusion, and reflecting to improve. Importantly, tutoring is not always action. Some of the most effective moments in tutoring are moments of \emph{restraint}: letting a learner struggle productively rather than interrupting with a hint \cite{kapur2008productive}. Human tutors do this naturally by prompting students to articulate what they know, when they are not sure, and how they would verify a step \cite{chi2001learning}. A content tutor says ``use $F=mg$'', while a metacognitive tutor asks: ``What are you trying to find?'' or ``What would convince you this step is correct?''

These teachable process skills fall under \textbf{metacognition} and \textbf{self-regulated learning (SRL)} \cite{flavell1979metacognition,zimmerman2002srl}. In educational psychology, metacognitive regulation comprises four key components: \textbf{Planning} (setting goals and selecting strategies before acting), \textbf{Monitoring} (tracking comprehension and progress during work and surfacing uncertainty), \textbf{Debugging} (repairing detected problems through alternative strategies or resources), and \textbf{Evaluation} (reflecting on progress and supporting transfer to new problems).

Most LLM-based tutoring systems implicitly optimize for \emph{helpfulness-as-output}: generating the next plausible hint or response under instructions (scaffolding, socratic questioning, mistake correction), or selecting among high-level pedagogical behaviors \cite{macina2025mathtutorbench,sonkar2023class,student-paradox}. This can obscure the core tutoring decision: \emph{what should a tutor do \underline{right now}, given the learner's signals and trajectory}. Progress is bottlenecked by the lack of (1) a shared, interpretable \emph{action space} for metacognitive tutoring; (2) scalable supervision with validity checks for those actions; (3) a benchmark that directly tests metacognitive tutoring as \emph{decision-making under pedagogy}.

MetaCLASS reframes ``AI tutoring'' from content delivery to \emph{coaching the learning process}. We operationalize these metacognitive processes as an interpretable \emph{action space} of coach moves (Table~\ref{tab:coach_moves}), treating \texttt{no\_intervention} as a first-class action to formalize restraint. Additionally, by turning metacognitive tutoring into a problem of \emph{pedagogical action selection}, we make an explicit coaching decision of which metacognitive move to take next, given the learner’s signals and trajectory.

In this paper, we make \textbf{three key contributions}:

\begin{enumerate}
    \item \textbf{MetaCLASS Framework.} 
    We introduce the first learning-science grounded framework for LLM-based metacognitive tutoring that targets self-regulated learning. We derive \textbf{11 interpretable coach moves} grounded in the Metacognitive Awareness Inventory (MAI), and map them systematically to LLMs' internal reasoning steps. The framework treats \texttt{no\_intervention} as a first-class pedagogical action.

    \item \textbf{MetaCLASS Dataset and Analysis.} We generate \textbf{1,015} metacognitive coaching conversations spanning \textbf{7,711} turns across GSM8K, MATH, and AIME datasets, annotated with turn-level coach move labels and learner metadata (calibration, help-seeking), alongside comprehensive dataset analyses.

    \item \textbf{Benchmark and Findings.} We introduce the \textbf{Coach Move Prediction} task which predicts the next metacognitive coach move given a problem and dialogue context. We observe a \textbf{compulsive intervention bias}, where models over-produce high-intervention moves and severely under-predict \texttt{no\_intervention}.

\end{enumerate}

\begin{table*}[t]
\centering
\begin{adjustbox}{max width=\textwidth}
\begin{tabular}{cccc}
\toprule
\textbf{MAI} & \textbf{Move} & \textbf{Function} & \textbf{Example Response} \\
\midrule
\multirow{2}{*}{\textbf{Planning}} 
& MONITOR\_GOAL (MG) & Elicit goal awareness & ``What are you trying to find?'' \\
& MONITOR\_PLAN (MP) & Elicit strategy awareness & ``What's your approach?'' \\
\midrule
\multirow{3}{*}{\textbf{Monitoring}}
& CHECK\_PROGRESS (CP) & Prompt progress tracking & ``How's it going?'' \\
& CHECK\_CONSISTENCY (CC) & Surface contradictions & ``Does that fit with what you said?'' \\
& UNCERTAINTY\_PROBE (UP) & Localize confusion & ``What's making you hesitate?'' \\
\midrule
\multirow{3}{*}{\textbf{Debugging}}
& STRATEGY\_ALTERNATIVE (SA) & Invite different approach & ``What else could you try?'' \\
& REPRESENTATION\_SHIFT (RS) & Suggest re-framing & ``Would a diagram help?'' \\
& PROMPT\_RESOURCE (PR) & Prompt help-seeking awareness & ``What would help you move forward?'' \\
\midrule
\textbf{Evaluation} & REFLECT\_OUTCOME (RO) & Prompt retrospective reflection & ``What worked here?'' \\
\midrule
\multirow{2}{*}{\textbf{Dialogue}}
& CONTINUE\_PROMPT (CT) & Neutral nudge & ``Keep going.'' \\
& NO\_INTERVENTION (NI) & Preserve productive flow & (silence) \\
\bottomrule
\end{tabular}
\end{adjustbox}
\caption{MetaCLASS Coach Move Taxonomy: 11 interpretable coach moves mapped to MAI regulatory processes. The moves support metacognitive regulation rather than directly delivering content.}
\label{tab:coach_moves}
\end{table*}

\section{MetaCLASS Framework}

In this section, we first ground the framework by mapping MAI theory to LLM reasoning patterns, deriving 11 interpretable coach moves. We then detail the two-phase architecture used to generate contingent, metacognitive coaching dialogues (Framework in Figure~\ref{fig:diagram}, Example in Section~\ref{app:example}).

\subsection{Connecting MAI to LLMs Reasoning}

The generation pipeline takes an LLM reasoning chain as input and produces a tutoring conversation with ground-truth coach move labels. We first review MAI's regulatory factors, then show how LLM reasoning chains map onto them, providing the principled basis for trajectory construction in the generation pipeline.

\subsubsection{Metacognitive Awareness Inventory}

MetaCLASS framework is grounded in the Metacognitive Awareness Inventory \cite{schraw1994assessing}, which characterizes how learners control their cognitive processes during learning through four empirically validated factors.

\begin{itemize}
    \item \textbf{Planning} includes pre-task activities, such as goal setting, strategy selection, and resource allocation (e.g., I set specific goals before I begin a task).

    \item \textbf{Monitoring} captures ongoing awareness of comprehension during learning, enabling detection of confusion, impasses, or inconsistencies (e.g., I find myself pausing regularly to check my comprehension).
    
    \item \textbf{Debugging} encompasses error correction strategies triggered when Monitoring detects problems (e.g., I re-evaluate my assumptions when I get confused).
    
    \item \textbf{Evaluation} involves post-task reflection on performance and strategy effectiveness, supporting transfer to future tasks (e.g., I ask myself how well I accomplished my goals once I'm finished).
\end{itemize}

These four factors form a validated taxonomy of metacognitive regulation, which MetaCLASS uses as a principled basis for designing coach moves.

\subsubsection{From LLMs Reasoning Chains to Coach Moves}
\label{sec:reasoning_to_moves}

MetaCLASS uses LLM reasoning chains as think-aloud scaffolds that capture the structured problem-solving process the student works through. To assign coach moves at each point in a conversation, we need this process to have a consistent structure. Recent work \cite{marjanovic2025deepseek} analyzes reasoning chains from DeepSeek-R1 \cite{guo2025deepseek} and identifies four consistent phases: \textbf{DEFINE} (reformulate the problem and identify goals), \textbf{BLOOM} (decompose and execute initial steps), \textbf{CYCLE} (reconsider assumptions and detect errors through self-monitoring), and \textbf{FINAL} (commit to an answer with confidence assessment).

These phases map systematically onto MAI's regulatory factors, providing the mechanism by which trajectories are constructed in the generation pipeline:

\begin{itemize}
    \item \textbf{DEFINE $\rightarrow$ Planning}: Both set goals before execution.
    \item \textbf{BLOOM $\rightarrow$ Planning}: Both decompose the problem and select a strategy.
    \item \textbf{CYCLE $\rightarrow$ Monitoring + Debugging}: Both involve self-monitoring and adjusting strategies when stuck.
    \item \textbf{FINAL $\rightarrow$ Evaluation}: Both assess performance and reflect on the outcome.
\end{itemize}

This mapping has three key implications. First, it grounds LLM reasoning in validated learning science. Second, it enables principled coach design: each MAI factor prescribes interventions for the corresponding phase. Third, it positions MetaCLASS coach moves as external scaffolding of LLMs' internal metacognitive processes.

\subsection{Metacognitive Coach Moves}

We operationalize MAI's four regulatory factors into eleven coach moves (Table~\ref{tab:coach_moves}). Each move translates an internal metacognitive process into an external prompt. Critically, these moves prompt reflection on the problem-solving process itself, which should never provide domain knowledge, evaluate correctness, or reveal solutions. A special coach move is \texttt{PROMPT\_RESOURCE} (PR), derived from help-seeking research~\cite{roll2011helpseeking}, which could give hints, but only when students initiate requests. Additionally, two dialogue moves support natural flow: \texttt{CONTINUE\_PROMPT} (CT) provides neutral nudges ("Keep going"), while \texttt{NO\_INTERVENTION} (NI) preserves productive struggle through silence.

\subsection{Two-Phase Generation Architecture}
\label{sec:planning}

MetaCLASS uses a two-phase architecture. Phase 1 instantiates a student model, analyzes the problem for learning obstacles, and integrates this information with the LLM reasoning chain and coach move taxonomy to construct a pedagogical trajectory. Phase 2 conditions dialogue generation on the student model and this trajectory, grounding every coach move in a planned pedagogical event.

\subsubsection{Phase 1: Pedagogical Planning}

\paragraph{Student Modeling.}
Effective coaching requires adapting to individual differences. MetaCLASS models students along two orthogonal dimensions (Calibration, Help-Seeking).

\textbf{Calibration} refers to the accuracy of self-assessment: whether students' perceived understanding matches their actual understanding~\cite{kruger1999unskilled}.

\begin{itemize}

    \item \textbf{Over-Confident} students overestimate their understanding, using confident language ("obviously," "just") and dismissing challenges without detecting errors.
    \item \textbf{Under-Confident} students doubt correct reasoning, hedging constantly ("maybe," "I'm not sure") and hesitating to commit to answers.
    \item \textbf{Well-Calibrated} students accurately assess understanding, articulating specific confusions ("I understand X but I'm stuck on Y").
\end{itemize}

\textbf{Help-Seeking} captures patterns of resource utilization during learning \cite{aleven2006helpseeking,roll2011helpseeking}.

\begin{itemize}

    \item \textbf{Avoidant} students resist asking for help even when stuck, and view help requests as failure and persist unproductively. They are primary targets for resource prompts.
    \item \textbf{Executive} students use resources strategically, attempting problems first and integrating help into understanding. They are the target pattern for self-regulated learning.
    \item \textbf{Dependent} students request help prematurely and broadly, seeking answers rather than understanding. They should not immediately receive resource prompts.

\end{itemize}

Crossing these dimensions yields nine theoretical profiles, of which \textbf{eight are valid} (over-confident + executive is contradictory). Each profile demands different coaching strategies.

\paragraph{Problem Analysis.}
Each problem is analyzed to identify potential learning obstacles. \textit{Knowledge Gaps} (missing domain knowledge) are addressed through content hints; \textit{Strategy Gaps} (knowledge present but cannot apply) are addressed through scaffolding hints; \textit{Monitoring Gaps} (unnoticed errors) and \textit{Execution Gaps} (careless mistakes) are addressed through prompts that highlight inconsistency or direct attention to verification without revealing answers. Both content and scaffolding hints are also pre-generated as problem-specific resources during planning but delivered only when students explicitly request them.

For each problem, we also generate metacognitive support structures aligned with the MAI framework (planning, monitoring, debugging, and evaluation support). This structured decomposition serves as the pedagogical foundation for trajectory construction.

\paragraph{Trajectory Construction.}
The core of planning is the trajectory, a sequence of pedagogical events that serves as the causal backbone of the conversation. Each event specifies a four-part chain:

\begin{itemize}

    \item \textbf{Event}: What happens in the problem-solving process (e.g., student encounters an ambiguity, gets stuck on a concept)

    \item \textbf{Signal}: What observable behavior the student produces (e.g., hedging language, long pause)

    \item \textbf{Move}: What coach intervention is appropriate given the student's profile (e.g., uncertainty probe, consistency check)

    \item \textbf{Effect}: What change should result from the intervention (e.g., student articulates confusion, requests help)

\end{itemize}

\subsubsection{Phase 2: Dialogue Generation}

This phase transforms the established LLM reasoning and trajectory into a coherent student-coach dialogue. The generation process strictly enforces the pedagogical plan: students exhibit behaviors consistent with their assigned profile, encounter the anticipated learning obstacles, and progress through the trajectory's event sequence.

\paragraph{Help System Design}

The help system models student-initiated resource use. Unlike traditional tutoring systems where hints are offered by the tutor, MetaCLASS generates hints during planning but delivers them only when students explicitly request help.

\textbf{Hint Types.} \textit{Content hints} provide domain knowledge (formulas, facts) for Knowledge Gaps, while \textit{scaffolding hints} provide process guidance (strategies, approaches) for Strategy Gaps.

\textbf{Help Quality.} We evaluate the appropriateness of help requests: \textit{appropriate} (asked when genuinely stuck, chose correct type), \textit{premature} (asked without real attempt), \textit{delayed} (asked too late after excessive struggle), or \textit{mismatched} (asked for wrong hint type).

\begin{table}[t]
\centering
\small
\resizebox{\linewidth}{!}{%
\begin{tabular}{ccccc}
\toprule
\textbf{Dataset} & \textbf{\# Conv} & \textbf{\# Turn} & \textbf{Contingent} & \textbf{Trajectory} \\
\midrule
GSM8K & 496 & 3,180 & 98.6\% & 99.4\% \\
MATH  & 490 & 4,142 & 96.2\% & 100\% \\
AIME  & 29  & 389   & 97.3\% & 100\% \\
\midrule
\textbf{Total / Avg} & 1,015 & 7,711 & 97.3\% & 99.8\% \\
\bottomrule
\end{tabular}%
}
\caption{MetaCLASS dataset statistics and evaluation results for contingent scaffolding and trajectory following.}
\label{tab:datasets}
\end{table}

\section{MetaCLASS Dataset}

\subsection{Dataset Statistics}

Following MetaCLASS framework, we generate student-coach conversations with GPT-5.1~\cite{achiam2023gpt} model, across three math datasets of varying difficulty: GSM8K (grade school math word problems) ~\cite{cobbe2021gsm8k}, MATH (high school competition)~\cite{hendrycks2021math}, and AIME (advanced olympiad)~\cite{maa2024aime}. As shown in Table~\ref{tab:datasets}, the dataset comprises 1,015 conversations totaling 7,711 turns. Examples of generated trajectories and conversations can be found in Appendix~\ref{app:example}.

\subsection{Dataset Validation}
\label{sec:dataset_validation}

We validate the MetaCLASS dataset through nine complementary analyses grounded in learning science research, examining whether generated conversations exhibit the structural and pedagogical properties of effective tutoring. We present three core analyses below; six additional analyses (MAI factor coverage, move sequence structure, student voice authenticity, help-seeking trajectories, internal field coherence, and post-hint coaching sequences) appear in Appendix~\ref{app:additional_validation}, along with detailed methodology and per-dataset breakdowns.

\paragraph{Contingent Scaffolding.}
Contingent scaffolding assesses whether the coach adapts to the learner's observed state rather than following a fixed script~\cite{wood1976role}. We identify difficulty signals while excluding productive exploration, then test whether the coach's next move is appropriate for the conversation phase (Appendix~\ref{app:contingent_method}). Across 1,890 signals in 828 conversations, coaches select appropriate moves 97.4\% of the time (Table~\ref{tab:datasets}), with 92--100\% rates across phases and datasets. Most remaining cases substitute one reasonable move for another rather than making pedagogically inappropriate choices.

\paragraph{Profile-Differentiated Coaching.}
A main claim of MetaCLASS is that the coach adapts to individual student profiles. We validate this by examining intervention timing, move selection, and student behavior across all three datasets.

\textit{Intervention Timing.} Table~\ref{tab:ttfi} shows the Time-to-First-Intervention (TTFI), the number of student turns before the coach's first active move. Across all datasets, avoidant students receive later intervention than dependent students. This differentiation respects Kapur's productive struggle principle~\cite{kapur2008productive}: avoidant students benefit from more struggle time to build help-seeking confidence, while dependent students benefit from earlier scaffolding.

\textit{External Help Usage.} Beyond coach behavior, student behavior also differs systematically by profile. Aggregating across datasets, Dependent students invoke external help (the \textsc{system} speaker) most frequently (97.1\% of conversations), Executive students at intermediate rates (80.0\%), and Avoidant students least often (63.1\%). The 34-point gap between Dependent and Avoidant students confirms that help-seeking profiles produce meaningfully different student behaviors, not just different coach responses.

\paragraph{Trajectory Alignment.}
We assess trajectory-execution fidelity by comparing planned and realized coach moves. As shown in Table~\ref{tab:datasets}, 99.8\% of conversations follow the planned trajectory. Across datasets, trajectories average 4.3 planned moves spanning 4.1 unique move types, indicating alignment with nontrivial pedagogical plans. The three misaligned GSM8K cases reflect early termination with \texttt{REFLECT\_OUTCOME} when students solved the problem sooner than expected. As a complementary check, two independent domain experts annotated 30 randomly sampled conversations (10 per dataset) and both agreed that all sampled conversations align with their planned trajectories.

\begin{table}[t]
\centering
\small
\begin{tabular}{lccc}
\toprule
\textbf{Help-Seeking} & \textbf{GSM8K} & \textbf{MATH} & \textbf{AIME} \\
\midrule
Avoidant   & 1.6--2.3 & 1.4--1.7 & 1.5--2.7 \\
Executive  & 1.1--1.5 & 1.1      & 1.0 \\
Dependent  & 1.1--1.3 & 1.1--1.2 & 1.0--1.2 \\
\bottomrule
\end{tabular}
\caption{Time-to-First-Intervention (TTFI) across calibration sub-profiles, grouped by help-seeking type and dataset.}
\vspace{-5pt}
\label{tab:ttfi}
\end{table}

\section{Coach Move Prediction Benchmark}
\subsection{Task Definition}

We define \textbf{Coach Move Prediction} as inference with a prompted, fixed LLM: given a math problem and the tutoring dialogue up to the current student turn, the model must select the next metacognitive coaching action. Let $x=(P,H_t)$ denote the input, where $P$ is the math problem and
$H_t = \{(u_1,m_1,r_1),\ldots,(u_{t-1},m_{t-1},r_{t-1}),u_t\}$ is the dialogue history up to the current student utterance $u_t$.
Here $u_i$ is the student utterance at turn $i$, $m_i$ is the coach move label, and $r_i$ is the coach response text.

The model outputs \emph{structured text} containing (i) a discrete coach move $m_t$ and (ii) an optional natural-language coach response $r_t$. We evaluate only the move decision: the predicted move $\hat m_t$ is extracted by parsing the model output and must belong to a finite action space $\mathcal{M}$ of 11 metacognitive coach moves (Table~\ref{tab:coach_moves}).

Conceptually, prompting and decoding induce an implicit distribution over moves, and we decode a single prediction:

\begin{equation}
\hat{m}_t = \arg\max_{m \in \mathcal{M}} P_{\theta}(m \mid x),
\end{equation}
where $P_{\theta}$ is the distribution of a fixed pretrained model under our prompt template and decoding procedure. We report accuracy of $\hat m_t$ against the ground-truth move $m_t$ specified by the MetaCLASS pedagogical trajectory. If the model outputs an invalid or unparsable label (not in $\mathcal{M}$), we count the prediction as incorrect.

\begin{table}[t]
\centering
\resizebox{\columnwidth}{!}{
\small
\setlength{\tabcolsep}{4pt}
\begin{tabular}{l|rr|rr|rr}
\toprule
& \multicolumn{2}{c|}{\textbf{GSM8K}} & \multicolumn{2}{c|}{\textbf{MATH}} & \multicolumn{2}{c}{\textbf{AIME}} \\
\textbf{Model} & Full & Min & Full & Min & Full & Min \\
\midrule
\rowcolor{gray!20}
\multicolumn{7}{l}{\textit{Large Models} ($>=$ 70B)} \\
\midrule
GPT-oss-120b & 36.2 & \textbf{32.9} & 30.8 & 27.1 & 25.2 & 19.3 \\
Qwen3-80b-Think & \textbf{43.2} & 32.3 & \textbf{37.5} & \textbf{27.6} & \textbf{29.6} & 20.3 \\
Qwen3-80b-Inst & 36.2 & 29.7 & 30.4 & 26.4 & 22.1 & \textbf{24.7} \\
LLaMA-70b & 36.0 & 29.3 & 30.1 & 23.0 & 22.4 & 18.3 \\
\midrule
\rowcolor{gray!20}
\multicolumn{7}{l}{\textit{Small Models} ($<=$ 20B)} \\
\midrule
GPT-oss-20b & \textbf{34.5} & \textbf{29.8} & \textbf{29.3} & \textbf{23.8} & 21.9 & \textbf{17.5} \\
Phi4-14b & 33.7 & 25.1 & 29.1 & 21.3 & 21.1 & 18.8 \\
LLaMA-8b & 28.5 & 20.8 & 25.9 & 17.9 & 15.7 & 14.4 \\
Qwen3-4b & 34.4 & 5.1 & 28.4 & 5.9 & \textbf{22.6} & 6.4 \\
Phi4-4b & 27.0 & 19.8 & 22.3 & 16.8 & 13.1 & 9.8 \\
\bottomrule
\end{tabular}
}
\caption{Model performance on Coach Move Prediction across three mathematical reasoning datasets using full prompt versus minimal prompts. Best accuracy per column is in \textbf{bold}.}
\label{tab:model_accuracy}
\vspace{-5pt}
\end{table}

\begin{table*}[t]
\centering
\resizebox{0.8\textwidth}{!}{
\begin{tabular}{l|rrrrrrrrrrr|r}
\toprule
\textbf{Model} & \textbf{MG} & \textbf{MP} & \textbf{CP} & \textbf{CC} & \textbf{UP} & \textbf{SA} & \textbf{RS} & \textbf{PR} & \textbf{RO} & \textbf{CT} & \textbf{NI} & \textbf{Avg} \\
\midrule
\rowcolor{gray!20}
\multicolumn{13}{l}{\textit{Large Models} ($>=$ 70B)} \\
\midrule
gpt-oss-120b & 23.0 & 60.5 & 33.1 & 31.7 & 69.6 & 23.8 & 33.0 & 9.0 & 98.6 & 15.4 & 4.5 & 32.7 \\
Qwen3-80b-Think & 10.1 & 54.4 & 5.4 & 33.6 & 67.2 & 12.6 & 7.2 & 8.7 & 98.1 & 0.0 & 29.0 & 39.5 \\
Qwen3-80b-Inst & 51.8 & 38.1 & 19.2 & 33.0 & 71.3 & 12.1 & 8.2 & 37.5 & 99.6 & 0.0 & 5.7 & 32.4 \\
LLaMA-70b & 11.3 & 63.4 & 19.7 & 16.7 & 71.5 & 34.4 & 7.2 & 23.3 & 99.7 & 7.7 & 6.9 & 32.1 \\
\midrule
\rowcolor{gray!20}
\multicolumn{13}{l}{\textit{Small Models} ($<=$ 20B)} \\
\midrule
gpt-oss-20b & 14.7 & 59.6 & 17.6 & 23.6 & 80.8 & 19.1 & 6.2 & 11.8 & 98.0 & 0.0 & 5.0 & 31.0 \\
Phi4-14b & 32.8 & 55.3 & 15.3 & 46.5 & 36.4 & 38.8 & 1.0 & 24.3 & 99.2 & 7.7 & 1.4 & 30.6 \\
LLaMA-8b & 18.1 & 46.8 & 23.2 & 28.6 & 64.9 & 22.1 & 3.1 & 15.3 & 79.9 & 7.7 & 2.0 & 26.4 \\
Qwen3-4b & 66.0 & 25.2 & 4.0 & 45.1 & 72.5 & 25.9 & 15.5 & 35.8 & 98.4 & 0.0 & 0.3 & 30.6 \\
Phi4-4b & 67.5 & 22.9 & 18.8 & 30.6 & 30.9 & 1.2 & 0.0 & 6.2 & 93.6 & 0.0 & 0.1 & 23.8 \\
\bottomrule
\end{tabular}
}
\caption{Per-move prediction accuracy (\%) for the 11 coach moves (Full prompt and aggregate across all datasets.)}
\label{tab:permove_accuracy}
\end{table*}

\subsection{Models and Evaluations}

\paragraph{Models.} We evaluate nine open-source LLMs spanning four families. Large models include GPT-oss-120B~\cite{agarwal2025gpt}, Qwen3-80B-Instruct, Think~\cite{yang2025qwen3}, and LLaMA-3.3-70B~\cite{touvron2023llama}. Small models include GPT-oss-20B, Qwen3-4B, Phi-4-14B~\cite{abdin2024phi}, Phi-4-4B, and LLaMA-3.1-8B. We focus on general-purpose LLMs because no existing model is purpose-built for metacognitive coaching with an explicit \texttt{NO\_INTERVENTION} action. We use hyperparameters following model card recommendations (see Appendix~\ref{app:parameters}).

\paragraph{Prompting Strategy.} We evaluate two prompt conditions (see Appendix~\ref{app:prompts}). The comparison tests a diagnostic hypothesis: do LLMs possess internalized metacognitive coaching ability, or do they require explicit instruction? \textit{Full prompt} provides (1) definitions and examples for all 11 coach moves, (2) move selection guidelines specifying when each is appropriate, and (3) coaching rules emphasizing metacognitive support over content. \textit{Minimal prompt} provides only move names without definitions, examples, or guidelines.


\begin{table}[t]
\centering
\resizebox{\columnwidth}{!}{
\setlength{\tabcolsep}{3pt}
\begin{tabular}{cccccccc}
\toprule
\textbf{Rank} & \textbf{Move} & \textbf{Acc (\%)} & \textbf{N} & \textbf{GT \%} & \textbf{Pred \%} & \textbf{Bias} \\
\midrule
1 & NI & 11.5 & 12,876 & 41.7 & 6.6 & \textcolor{red}{-35.2} \\
2 & SA & 20.7 & 1,360 & 4.4 & 2.9 & \textcolor{red}{-1.5} \\
3 & PR & 19.6 & 1,152 & 3.7 & 2.5 & \textcolor{red}{-1.2} \\
4 & MG & 24.1 & 1,304 & 4.2 & 4.0 & \textcolor{red}{-0.2} \\
5 & CC & 28.7 & 3,192 & 10.3 & 10.9 & \textcolor{blue}{+0.6} \\
6 & RS & 13.9 & 388 & 1.3 & 2.0 & \textcolor{blue}{+0.7} \\
7 & CP & 19.4 & 1,704 & 5.5 & 6.4 & \textcolor{blue}{+0.9} \\
8 & CT & 5.8 & 52 & 0.2 & 1.8 & \textcolor{blue}{+1.6} \\
9 & MP & 54.1 & 2,632 & 8.5 & 15.2 & \textcolor{blue}{+6.7} \\
10 & UP & 69.9 & 2,120 & 6.9 & 16.0 & \textcolor{blue}{+9.1} \\
11 & RO & 99.0 & 4,064 & 13.2 & 31.5 & \textcolor{blue}{+18.4} \\
\bottomrule
\end{tabular}
}
\caption{Move-level accuracy and prediction bias (large models with full prompt for all datasets). Ranked by prediction bias from most under-predicted to most over-predicted, with N showing ground truth instances. Bias (Pred \% - GT \%) reveals substantial under-prediction of restraint (NI: -35.2) and over-prediction of high-intervention moves (RO: +18.4).}
\label{tab:move_bias}
\vspace{-10pt}
\end{table}

\paragraph{Evaluation Metrics.} Our primary evaluation is \textit{Accuracy}, reported both overall and per-move. We also report \textit{NI Detection rate} for assessing whether LLMs can support students with appropriate silent restraints.

\begin{figure*}[t]
\centering
\begin{minipage}{0.48\textwidth}
    \centering
    \includegraphics[width=\textwidth]{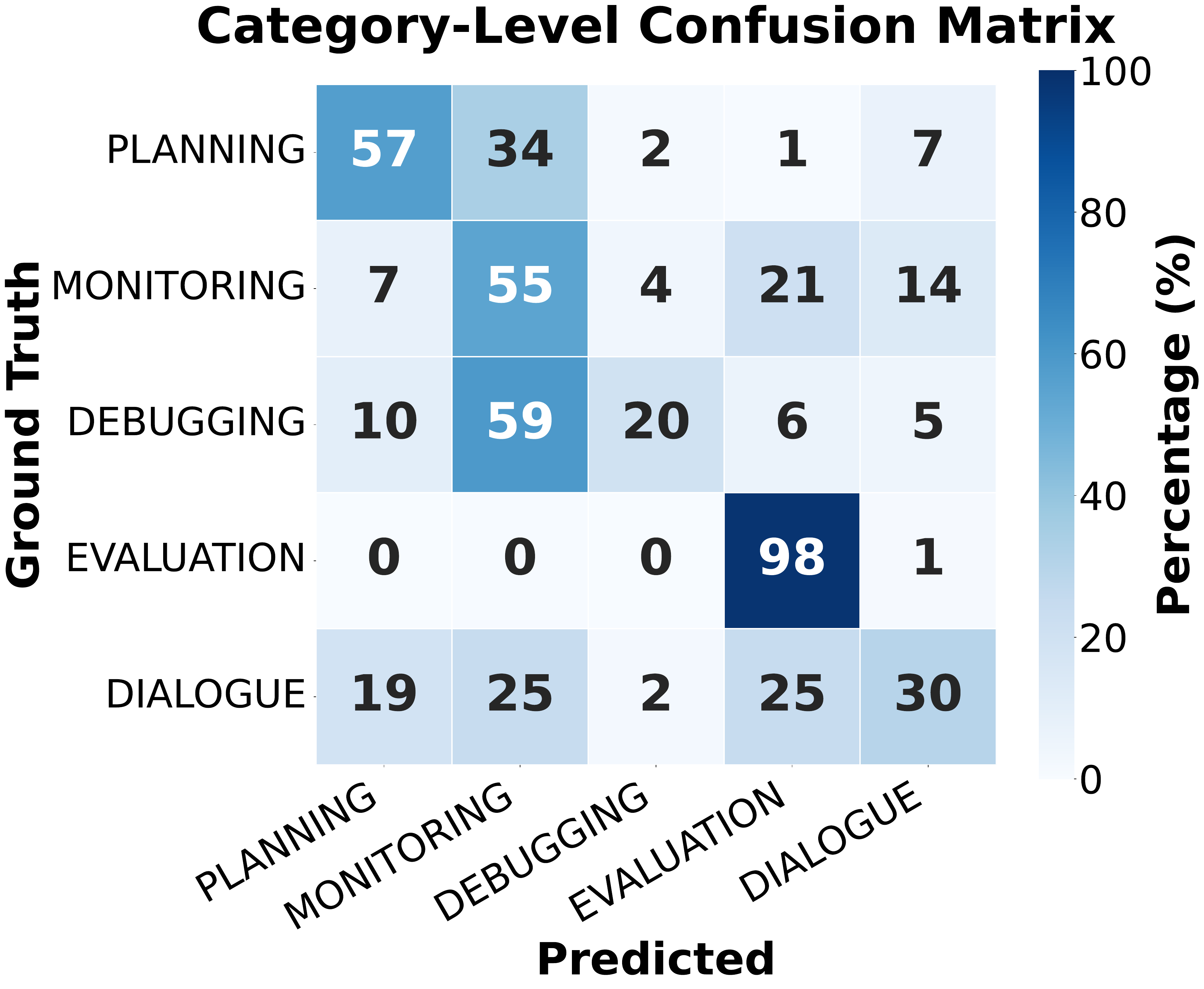}
    \caption{Category-level confusion matrix for Qwen3-80b-Think. DEBUGGING moves are predominantly misclassified as MONITORING (59\%), suggesting the model struggles to distinguish exploratory probing from corrective intervention.}
    \label{fig:category_cm}
\end{minipage}
\hfill
\begin{minipage}{0.48\textwidth}
    \centering
    \includegraphics[width=\textwidth]{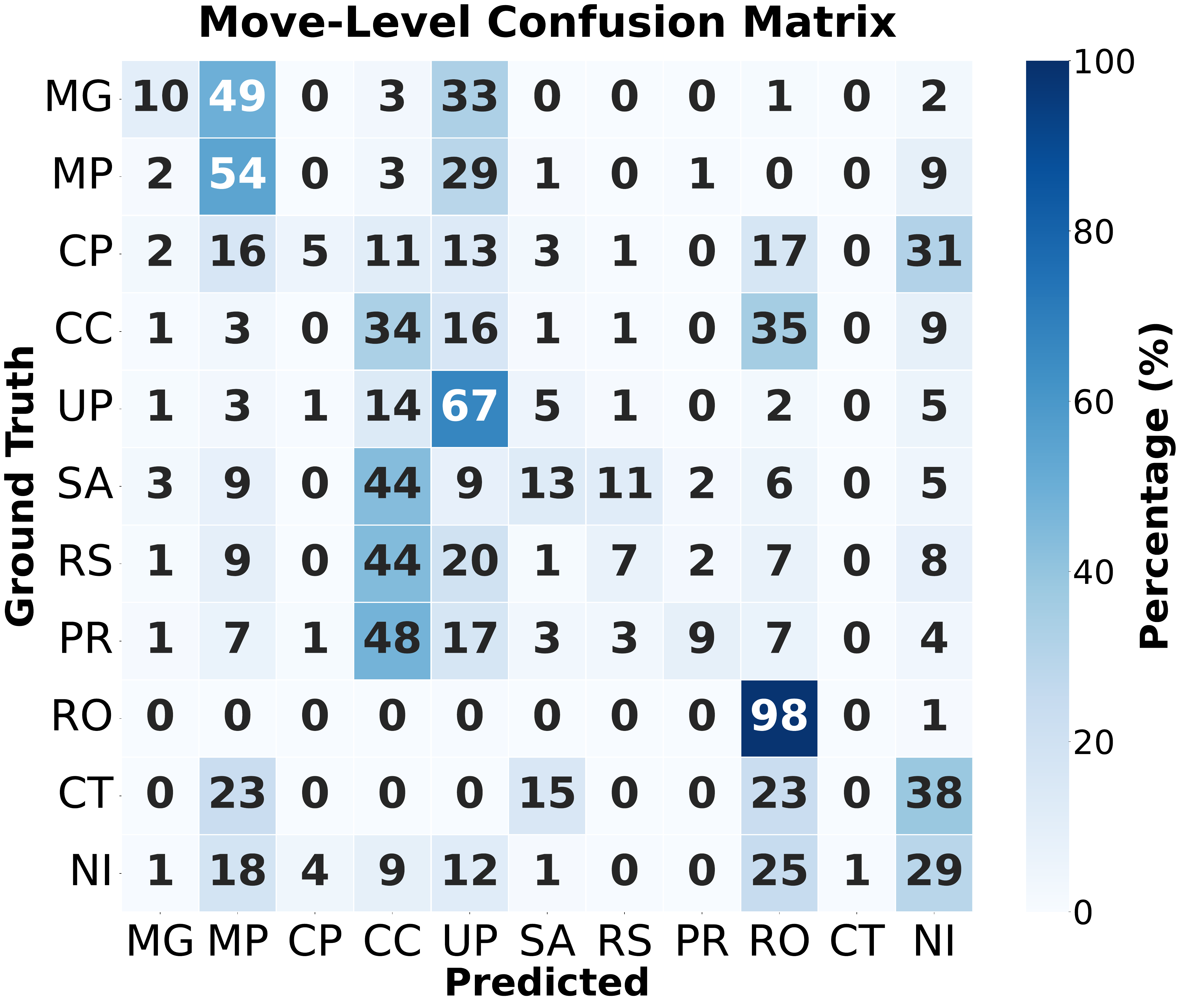}
    \caption{Move-level confusion matrix for Qwen3-80b-Think. \textsc{NI} reaches only 29\% accuracy, despite comprising 42\% of ground truth, revealing compulsive intervention bias.}
    \label{fig:move_cm}
\end{minipage}
\end{figure*}

\section{Results and Discussion}

Our analysis reveals three fundamental failures distinguishing metacognitive coaching from regular tutoring: (1) models generally struggle with metacognitive move prediction; (2) models exhibit compulsive intervention bias that explicit prompt instruction cannot override; and (3) models tend to favor consistency checks over strategic redirection. 

\subsection{Performance Across Models and Prompts}

Table~\ref{tab:model_accuracy} shows overall accuracy for coach move prediction. Model accuracy plateaus at 43.2\%, revealing that open-source LLMs still struggle to identify when and how to intervene ``metacognitively''. This contrasts with their strong performance on content-based tutoring~\cite{scarlatos2025training, sonkar2023class}, showing that knowing how to explain does not translate to knowing when to intervene. While large models generally achieve higher accuracy, smaller models are inconsistent. For instance, Qwen3-4B occasionally matches larger models yet exhibits catastrophic drops under certain configurations (5.1\% on GSM8K with Minimal prompt).

\paragraph{Impact of Context (Full vs Minimal).}
Full prompts with explicit MAI definitions consistently outperform Minimal prompts. Qwen3-4B shows this dependency most starkly, with the performance dropping from 34.4\% to 5.1\% without explicit instruction. This reveals LLMs lack internalized metacognitive tutoring ability and cannot autonomously apply coaching strategies without explicit prompting.

\subsection{Move-level Analysis}

\paragraph{Accuracy Variation.} 
Accuracy varies dramatically across moves (Table~\ref{tab:permove_accuracy}): near-perfect on \texttt{RO} ($\sim$99\%) versus near-zero on \texttt{NI} ($\leq$29.0\%) and \texttt{CT} ($\leq$15.4\%). High-accuracy moves follow clear surface patterns (\texttt{RO} at completion, \texttt{UP} after hesitation, \texttt{MP} after problem statements). Mid-accuracy moves (\texttt{CC} $\leq$59.5\%, \texttt{MP} $\leq$68.0\%) show high model-to-model variance, suggesting some LLMs capture these relevant patterns.

\paragraph{The Compulsive Intervention Bias.}
The most critical finding is models' systematic inability to withhold intervention. As shown in Table~\ref{tab:move_bias}, \texttt{NI}, the skill of recognizing when to remain silent, proves to be the hardest frequent move, with accuracy frequently below 7\% despite being the most common ground truth label (41.7\%). This occurs even though prompts explicitly specify \textit{"NI should be $\sim$35-50\% of coach turns."} Models exhibit average \texttt{NI} accuracy of 11.5\% with massive negative prediction bias of $-$35.2\%, dramatically under-predicting silence. This pattern suggests instruction-tuning of LLMs creates deeply ingrained intervention patterns that prompting cannot override. Even thinking-enabled models, which achieve 5$\times$ better \texttt{NI} detection than instruction-tuned counterparts (29.0\% vs. 5.7\%), still intervene in over 70\% of cases where restraint is appropriate. Conversely, models over-predict high-intervention moves like \texttt{RO} with significant positive bias (+18.4\%) and near-perfect accuracy (99\%).

\subsection{Failure Mode Analysis}
To understand systematic misprediction patterns, we analyzed confusion matrices for Qwen3-80B-Think in Figures~\ref{fig:category_cm} and \ref{fig:move_cm}.

\paragraph{Planning-Monitoring conflation.} 
\texttt{MG} is frequently confused with \texttt{MP} (49\%), with both exhibiting high cross-predictions with \texttt{UP} (33\% and 29\% respectively). This suggests models cannot distinguish goal-setting from strategy-setting or from localizing confusions.

\paragraph{Strategy redirection misdiagnosed as verification.} 
\texttt{SA} and \texttt{RS} show low accuracy with 44\% confusion with \texttt{CC}. When students need alternative approaches, models favor consistency checks over strategic redirection. Figure~\ref{fig:category_cm} confirms this: 59\% of Debugging moves are mispredicted as Monitoring, showing models recognize stuck students but systematically choose verification over strategy change.

\paragraph{Resource prompting nearly invisible.}
\texttt{PR} achieves only 9\% accuracy, most often substituted with \texttt{CC} (48\%). Models cannot distinguish students needing verification support from those requiring help-seeking scaffolding, leaving help-avoidant students without appropriate intervention.

\subsection{Implications for AI Tutoring Systems}

These findings have critical implications for LLM-based tutoring systems. Current models produce 6-10 times more interventions than is appropriate, training students to wait for hints rather than develop transferable metacognitive skills like planning, monitoring, and debugging. Because this bias stems from instruction-tuning's reward structure~\cite{dinucu2025problem}, the same over-intervention pattern can extend to tutoring beyond the math domain. Additionally, this compulsive intervention bias is invisible in standard evaluation. Metrics assessing helpfulness or satisfaction cannot detect over-intervention, and students may even prefer excessive scaffolding despite its harm to long-term development. Without explicit ground truth for restraint, as MetaCLASS provides, this pedagogical failure remains hidden behind metrics that reward responsiveness over developmental appropriateness.

\section{Related Work}

\subsection{Metacognition/ Self-Regulated Learning}

Metacognition concerns learners' awareness and control over their cognition \cite{flavell1979metacognition}: in SRL, learners set goals, select strategies, monitor understanding, repair breakdowns, and evaluate outcomes \cite{zimmerman2002srl}. Effective tutoring must therefore support learners' \emph{regulation} of problem-solving, not just deliver content. However, SRL constructs such as ``monitoring'' lack operational definitions at the granularity of turn-level tutoring decisions. MetaCLASS closes this gap through regulation processes, help-seeking, and calibration.

\paragraph{Help-seeking as coached self-regulation.}
Help-seeking is itself a metacognitive skill: learners must decide \emph{when} help is needed, \emph{what} kind to request, and \emph{how} to use it \cite{aleven2006helpseeking,roll2011helpseeking}.

\paragraph{Calibration and confidence signals.}
Calibration, the accuracy of learners' self-assessments \cite{kruger1999unskilled}, provides observable signals for coaching. Over-confident learners may miss errors; under-confident learners may abandon correct reasoning. Well-calibrated learners localize uncertainty (``I understand X but I'm stuck on Y''), enabling targeted monitoring and debugging interventions.

\paragraph{Restraint as a metacognitive tutoring decision.}
Metacognitive support does not require constant intervention. Human tutors use selective prompting and timing \cite{chi2001learning}, and productive failure research shows that initial struggle benefits learning \cite{kapur2008productive}. MetaCLASS therefore treats ``no intervention" as a meaningful tutoring action and explicit benchmark label.

\subsection{Intelligent Tutors}

Classic Intelligent Tutoring Systems (ITS) \cite{graesser2004autotutor,anderson1995cognitive} demonstrated that adaptive feedback improves learning, but adaptation in these systems is typically operationalized via correctness tracing rather than explicit metacognitive regulation. Recent LLM tutors excel at explanation generation \cite{macina2025mathtutorbench,sonkar2023class}, yet tend to prioritize \emph{utterance quality} (helpfulness, naturalness) over \emph{pedagogical decision-making}. At the content level, \cite{wang2024bridging} models remediation by identifying student math errors and selecting correction strategies, and MalRuleLib \cite{malrulelib} provides executable misconception rules with step traces for modeling student mathematical thinking. At the discourse level, talk move prediction \cite{ganesh2021would,suresh2022talkmoves,cao2025enhancing} classifies teacher and student moves in classroom settings using Accountable Talk theory.

What remains underexplored across these work is an interpretable, theory-grounded metacognitive action space that separates \emph{process coaching} from \emph{content delivery} while also recognizing \textit{no intervention} as a meaningful pedagogical action. MetaCLASS fills this gap by formalizing metacognition as a decision process: given learner signals and trajectory, the coach selects the next metacognitive move, including silence when non-intervention is pedagogically appropriate, thereby helping learners regulate their own thinking rather than primarily receiving answers. Our moves are grounded in MAI regulation theory rather than classroom discourse theory, and \textit{no intervention} is treated as a first-class pedagogical action. To our knowledge, MetaCLASS is the first framework to explicitly formalize when \textit{not} to intervene as part of tutoring competence.

\section{Conclusion and Future Work}

We presented \textbf{MetaCLASS}, a learning-science grounded framework and benchmark for \emph{metacognitive tutoring}. MetaCLASS operationalizes metacognitive regulation into 11 interpretable coach moves spanning planning, monitoring, debugging, and evaluation. To provide scalable, auditable supervision, we utilize a two-phase generation procedure that first plans pedagogical trajectories conditioned on learner profiles, then generates dialogue consistent with that plan, yielding a dataset with 7,711 conversation turns and explicit turn-level metacognitive labels and validation analyses testing contingency and trajectory following. Our Coach Move Prediction benchmark reveals that current LLMs struggle with pedagogical decision-making: the best model achieves only 43.2\% accuracy, and models consistently exhibit compulsive intervention bias by over-predicting high-intervention moves while severely under-predicting restraint (no intervention). These results suggest standard instruction-following tuning does not produce metacognitive tutoring competence.

Looking ahead, our findings suggest several directions. First, fine-tuning on MetaCLASS trajectories and RL-based alignment may improve restraint and reduce unnecessary intervention. Second, NI rate may serve as a practical diagnostic for over-intervention beyond standard helpfulness metrics. Third, because compulsive intervention bias and MAI regulation are largely domain-independent, MetaCLASS may transfer beyond math to domains such as programming and science tutoring.

\section*{Limitations}

MetaCLASS is intended as a \emph{diagnostic benchmark} for metacognitive tutoring decisions: we operationalize metacognitive support as move selection in text dialogue and study it on math problem-solving settings (GSM8K/MATH/AIME). This focus prioritizes interpretability and controlled evaluation over broad coverage of domains and modalities. In addition, our benchmark uses a single target move per turn derived from an explicit pedagogical trajectory; while this enables consistent supervision and analysis, some contexts may admit multiple pedagogically reasonable moves that are not credited under strict matching. Accordingly, our results should be interpreted as evidence about current LLMs' ability to follow theory-grounded metacognitive intervention policies under a controlled setting, rather than as measurements of downstream student learning outcomes, which would require randomized controlled trials with real learners and is left to future work. Additionally, MetaCLASS conversations are LLM-generated rather than collected from live tutoring sessions. While our validation analyses (Section~\ref{sec:dataset_validation}, Appendix~\ref{app:additional_validation}) demonstrate that the generated data captures key behavioral and pedagogical properties grounded in learning science, validating the framework with real learners and tutors remains an important direction for future work.

\section*{Acknowledgments}
This work was supported by NSF SafeInsights, cooperative agreement 2153481, and a Google Cloud Computing Award.

\bibliography{references}

\appendix

\clearpage
\section{MetaCLASS Design Principles Grounded in Learning Science}
\label{sec:design_principles}

MetaCLASS translates findings from tutoring and SRL into \emph{generation constraints} and evaluation targets. These principles govern \emph{when} support should be offered, \emph{how much} to offer, and when the best move is to withhold help.

\paragraph{Contingent scaffolding (adaptive support).}
Tutoring effectiveness depends on providing support that is contingent on the learner’s observed state rather than a fixed script. Classic scaffolding work characterizes effective tutoring as dynamic assistance responsive to learner needs \cite{wood1976role}. A synthesis emphasizes three defining features of scaffolding: \emph{contingency}, \emph{fading}, and \emph{transfer of responsibility} \cite{vandePol2010scaffolding}. MetaCLASS enforces contingency by tying each coach move to an explicit \textit{Event$\rightarrow$Signal$\rightarrow$Move$\rightarrow$Effect} trajectory, ensuring interventions are triggered by learner signals rather than tutor preference.

\paragraph{Productive struggle and timing of instruction.}
Learning can benefit when students attempt sensemaking before receiving explanations. Productive Failure shows that attempting solutions (and even failing) can prepare learners to learn more deeply from subsequent instruction \cite{kapur2008productive}. Related work argues there is a ``time for telling'': direct explanation is more effective after learners have constructed differentiated knowledge structures through prior activity \cite{schwartz1998time}. In MetaCLASS, this motivates \texttt{no\_intervention} and low-directiveness moves early in a struggle episode, preserving space for learner-generated progress before steering.

\paragraph{Restraint and the assistance dilemma.}
Tutors face the \emph{assistance dilemma}: too little help leads to floundering, while too much help can reduce effort, shallow processing, and self-regulation \cite{koedinger2007assistance}. Analyses of human tutoring similarly show that effective tutors elicit student reasoning and selectively intervene \cite{chi2001learning}. MetaCLASS encodes restraint with an explicit target distribution (35--50\% \texttt{no\_intervention}) and by treating minimal dialogue acts as first-class pedagogical actions rather than failures to respond.

\paragraph{Prioritizing internal regulation before external resources.}
Support should diminish and responsibility should shift toward the learner as problem solving progresses. Work on faded guidance (e.g., fading worked solution steps) supports the idea that gradually removing scaffolds can promote independent problem solving and deeper engagement \cite{renkl2004fading}. MetaCLASS reflects this principle by prioritizing internal regulation moves (e.g., \texttt{strategy\_alternative}, \texttt{representation\_shift}) before external resource prompting, discouraging premature outsourcing when the learner can still make productive progress.

\paragraph{Strategic help-seeking as a target competency.}
Help-seeking is a central SRL behavior: learners must decide \emph{when} to seek help, \emph{what} help to request, and \emph{how} to use it \cite{aleven2006helpseeking}. Empirical ITS work demonstrates that metacognitive feedback can improve learners’ help-seeking actions and their use of assistance \cite{roll2011helpseeking}, and broader reviews summarize both the promise and limits of help features in ITSs \cite{aleven2016help}. MetaCLASS generates hints during planning but delivers them only upon student request; additionally, \texttt{prompt\_resource} is constrained to occur \emph{after} internal strategy moves, promoting disciplined, student-initiated resource use.

Together, these principles ground MetaCLASS in learning science: coaching is contingent, preserves productive struggle, avoids over-assistance, and treats help-seeking as a coached skill rather than a shortcut.
\label{sec:appendix}

\section{Hyperparameters}
\label{app:parameters}
Following model card recommendations, we use the following sampling parameters: for Qwen3 models in thinking mode, temperature 0.6, top-$p$ 0.95, and top-$k$ 20; for Qwen3 in non-thinking mode, temperature 0.7 and top-$p$ 0.8; and for gpt-oss models, temperature 1.0 and top-$p$ 1.0.

\section{Additional Dataset Validation}
\label{app:additional_validation}

This appendix provides detailed methodology and breakdowns supporting the core analyses in Section~\ref{sec:dataset_validation}, followed by six additional analyses that further validate the MetaCLASS dataset.

\subsection{Contingent Scaffolding: Methodology}
\label{app:contingent_method}

\paragraph{Signal Detection.} We detect two categories of student difficulty signals through keyword pattern matching:
\begin{itemize}[leftmargin=*, itemsep=2pt, topsep=2pt, parsep=0pt, partopsep=0pt]
    \item \textbf{Uncertainty signals}: expressions of doubt or hesitation, e.g., ``I'm not sure,'' ``maybe,'' ``confused,'' ``I don't know,'' ``uncertain,'' ``wait,'' etc.
    \item \textbf{Stuck signals}: expressions of impasse, e.g., ``stuck,'' ``can't figure,'' ``no idea,'' ``give up,'' ``this is hard,'' etc.
\end{itemize}

We filter out productive exploration (e.g., ``maybe I could try a different approach...''), which indicates active problem-solving rather than genuine confusion. Across datasets, this filtering removes 502 instances (GSM8K: 207, MATH: 278, AIME: 17), leaving 1,890 genuine difficulty signals in 828 conversations.

\paragraph{Phase-Appropriate Moves.} Conversations are divided into three phases based on normalized turn position. Different coach moves are considered appropriate responses to difficulty signals at each phase:
\begin{itemize}[leftmargin=*, itemsep=2pt, topsep=2pt, parsep=0pt, partopsep=0pt]
    \item \textbf{Early (0--25\%)}: \texttt{MG}, \texttt{MP}, \texttt{UP}, \texttt{CC}, \texttt{NI} (goal-setting and initial diagnosis).
    \item \textbf{Middle (25--75\%)}: \texttt{UP}, \texttt{CC}, \texttt{SA}, \texttt{RS}, \texttt{PR}, \texttt{CP}, \texttt{NI} (working and debugging).
    \item \textbf{Late (75--100\%)}: \texttt{UP}, \texttt{CC}, \texttt{RO}, \texttt{PR}, \texttt{NI} (verification and reflection).
\end{itemize}

\paragraph{Phase Breakdown.} Table~\ref{tab:contingent_phase} shows contingency rates by conversation phase for each dataset. Contingency holds uniformly across early, middle, and late phases on all three datasets, indicating that the coach adapts to learner difficulty throughout the conversation rather than only at particular stages.

\begin{table*}[t]
\centering
\small
\resizebox{0.8\linewidth}{!}{%
\begin{tabular}{l rrr rrr rrr}
\toprule
& \multicolumn{3}{c}{\textbf{GSM8K}} & \multicolumn{3}{c}{\textbf{MATH}} & \multicolumn{3}{c}{\textbf{AIME}} \\
\cmidrule(lr){2-4} \cmidrule(lr){5-7} \cmidrule(lr){8-10}
\textbf{Phase} & \textbf{App.} & \textbf{Tot.} & \textbf{Rate} & \textbf{App.} & \textbf{Tot.} & \textbf{Rate} & \textbf{App.} & \textbf{Tot.} & \textbf{Rate} \\
\midrule
Early & 285 & 287 & 99.3\% & 338 & 353 & 95.8\% & 23 & 25 & 92.0\% \\
Middle & 540 & 550 & 98.2\% & 496 & 511 & 97.1\% & 45 & 45 & 100\% \\
Late & 47 & 47 & 100\% & 62 & 67 & 92.5\% & 5 & 5 & 100\% \\
\midrule
\textbf{Overall} & \textbf{872} & \textbf{884} & \textbf{98.6\%} & \textbf{896} & \textbf{931} & \textbf{96.2\%} & \textbf{73} & \textbf{75} & \textbf{97.3\%} \\
\bottomrule
\end{tabular}
}
\caption{Contingent scaffolding rates by conversation phase and dataset. App. = appropriate responses, Tot. = total difficulty signals.}
\label{tab:contingent_phase}
\end{table*}

\subsection{MAI Factor Coverage}
\label{app:mai_coverage}

We measure whether each MAI regulation factor~\cite{schraw1994assessing} appears at least once per conversation. Aggregating across datasets, coverage rates are: Planning (80.9\%), Monitoring (97.5\%), Debugging (37.4\%), and Evaluation (100\%). The moderate Debugging rate is appropriate since debugging requires errors, which not all conversations contain. Critically, Debugging exhibits strong profile differentiation: Avoidant students engage in debugging 81.4\% of the time compared to only 9.7\% for Executive and 12.3\% for Dependent students. This is theoretically grounded, as Avoidant students struggle longer before seeking help and thus encounter more errors requiring strategy changes.

\subsubsection{Detailed Breakdowns}
\label{app:mai_method}

\paragraph{Move-to-Factor Mapping.} Each of the 11 coach moves maps to one of four MAI regulation factors~\cite{schraw1994assessing}, as shown in Table~\ref{tab:coach_moves}:

\begin{itemize}[leftmargin=*, itemsep=2pt, topsep=2pt, parsep=0pt, partopsep=0pt]
    \item \textbf{Planning}: \texttt{monitor\_goal}, \texttt{monitor\_plan}
    \item \textbf{Monitoring}: \texttt{check\_progress}, \texttt{check\_consistency}, \texttt{uncertainty\_probe}
    \item \textbf{Debugging}: \texttt{strategy\_alternative}, \texttt{representation\_shift}, \texttt{prompt\_resource}
    \item \textbf{Evaluation}: \texttt{reflect\_outcome}
\end{itemize}

A factor is marked as present in a conversation if at least one corresponding move appears.

\paragraph{Per-Dataset Rates.} Table~\ref{tab:mai_dataset} shows factor presence rates for each dataset individually.

\begin{table}[h]
\centering
\small
\resizebox{\columnwidth}{!}{%
\begin{tabular}{l rrrr}
\toprule
\textbf{Dataset} & \textbf{Planning} & \textbf{Monitoring} & \textbf{Debugging} & \textbf{Evaluation} \\
\midrule
GSM8K & 75.6\% & 95.0\% & 30.0\% & 100\% \\
MATH & 86.3\% & 99.8\% & 42.4\% & 100\% \\
AIME & 79.3\% & 100\% & 79.3\% & 100\% \\
\bottomrule
\end{tabular}
}
\caption{MAI factor presence rates by dataset. Debugging increases with problem difficulty (30.0\% $\rightarrow$ 79.3\%), reflecting that harder problems produce more errors requiring strategy changes.}
\label{tab:mai_dataset}
\end{table}

Debugging increases sharply with problem difficulty (30.0\% $\rightarrow$ 42.4\% $\rightarrow$ 79.3\%), reflecting that harder problems produce more errors requiring strategy changes. Planning and Monitoring remain stable across datasets, indicating that the coach consistently engages these factors regardless of difficulty.

\paragraph{Factor Rates by Calibration.} Table~\ref{tab:mai_calibration} shows how MAI factor coverage varies by calibration profile across datasets.

\begin{table}[h]
\centering
\small
\resizebox{\columnwidth}{!}{%
\begin{tabular}{ll rrrr}
\toprule
\textbf{Dataset} & \textbf{Calibration} & \textbf{Plan.} & \textbf{Mon.} & \textbf{Debug.} & \textbf{Eval.} \\
\midrule
\multirow{3}{*}{GSM8K} & Over & 63\% & 98\% & 23\% & 100\% \\
& Under & 89\% & 97\% & 36\% & 100\% \\
& Accurate & 70\% & 91\% & 29\% & 100\% \\
\midrule
\multirow{3}{*}{MATH} & Over & 76\% & 100\% & 42\% & 100\% \\
& Under & 92\% & 100\% & 47\% & 100\% \\
& Accurate & 88\% & 99\% & 39\% & 100\% \\
\midrule
\multirow{3}{*}{AIME} & Over & 67\% & 100\% & 67\% & 100\% \\
& Under & 100\% & 100\% & 73\% & 100\% \\
& Accurate & 67\% & 100\% & 92\% & 100\% \\
\bottomrule
\end{tabular}
}
\caption{MAI factor presence rates by calibration profile and dataset.}
\label{tab:mai_calibration}
\end{table}

Under-calibrated students consistently show the highest Planning rates (89--100\%), likely because their low confidence leads to more deliberate goal-setting. Debugging varies more with problem difficulty than with calibration profile, suggesting that error occurrence (rather than self-perception) drives debugging engagement.

\subsection{Profile-Differentiated Move Selection}
\label{app:move_selection}

\textit{Move Selection.} Coach move distributions differ systematically by student profile (Table~\ref{tab:move_by_profile}). Avoidant students receive \texttt{prompt\_resource} (7--10\%), encouraging help-seeking, while Dependent students receive zero \texttt{prompt\_resource} across all datasets but elevated \texttt{check\_consistency} (13--14\%), building self-verification skills rather than reinforcing dependency. Dependent students also receive fewer \texttt{uncertainty\_probe}s (3--6\% vs.\ 8--10\% for Avoidant), consistent with their tendency to seek help before engaging deeply enough to exhibit uncertainty. By calibration, Over-confident students receive elevated \texttt{check\_consistency} (18\%) compared to Under-confident or Accurate students (6--12\%) and fewer \texttt{uncertainty\_probe}s (3--7\% vs.\ 7--10\% for Under-confident), reflecting the Dunning-Kruger effect~\cite{kruger1999unskilled}: overconfident students need external verification prompts because they miss errors internally, while underconfident students' frequent hedging triggers more uncertainty probes.

\begin{table*}[ht]
\centering
\small
\resizebox{0.75\linewidth}{!}{
\begin{tabular}{l rrr rrr rrr}
\toprule
& \multicolumn{3}{c}{\textbf{GSM8K}} & \multicolumn{3}{c}{\textbf{MATH}} & \multicolumn{3}{c}{\textbf{AIME}} \\
\cmidrule(lr){2-4} \cmidrule(lr){5-7} \cmidrule(lr){8-10}
\textbf{Profile} & \textbf{UP} & \textbf{PR} & \textbf{CC} & \textbf{UP} & \textbf{PR} & \textbf{CC} & \textbf{UP} & \textbf{PR} & \textbf{CC} \\
\midrule
\rowcolor{gray!15}
\multicolumn{10}{l}{\textit{By Help-Seeking}} \\
\quad Avoidant & 10\% & 10\% & 7\% & 8\% & 8\% & 10\% & 9\% & 7\% & 12\% \\
\quad Executive & 10\% & 0\% & 6\% & 6\% & 0\% & 10\% & 9\% & 2\% & 12\% \\
\quad Dependent & 3\% & 0\% & 13\% & 3\% & 0\% & 14\% & 6\% & 0\% & 14\% \\
\midrule
\rowcolor{gray!15}
\multicolumn{10}{l}{\textit{By Calibration}} \\
\quad Over & 3\% & 3\% & 18\% & 5\% & 3\% & 18\% & 7\% & 4\% & 18\% \\
\quad Under & 10\% & 5\% & 6\% & 7\% & 4\% & 9\% & 9\% & 2\% & 11\% \\
\quad Accurate & 8\% & 4\% & 6\% & 6\% & 3\% & 10\% & 7\% & 4\% & 12\% \\
\bottomrule
\end{tabular}
}
\caption{Key coach move percentages by student profile across datasets. UP = \texttt{uncertainty\_probe}, PR = \texttt{prompt\_resource}, CC = \texttt{check\_consistency}. Avoidant students receive more resource prompts; Over-confident students receive elevated consistency checks; Dependent and Over-confident students receive fewer uncertainty probes.}
\label{tab:move_by_profile}
\vspace{-10pt}
\end{table*}

\subsection{Move Sequence Structure}
\label{app:move_sequence}

Effective tutoring follows structured pedagogical patterns, not random move selection~\cite{chi2001learning,graesser1995collaborative}. We analyze move positions within conversations and find a consistent temporal ordering across all three datasets (Table~\ref{tab:move_position}): goal-setting moves appear early (\texttt{monitor\_goal} at 0.01--0.08, \texttt{monitor\_plan} at 0.07--0.13), working moves cluster mid-conversation (\texttt{uncertainty\_probe} at 0.38--0.41, \texttt{check\_consistency} at 0.50--0.64), external help comes later (\texttt{prompt\_resource} at 0.70--0.72), and reflection closes the conversation (\texttt{reflect\_outcome} at 0.99). This sequence mirrors Graesser et al.'s five-step tutoring frame~\cite{graesser1995collaborative}: establish goal $\rightarrow$ diagnose understanding $\rightarrow$ work through problem $\rightarrow$ provide scaffolding $\rightarrow$ assess outcome.  Notably, the escalation chain \texttt{uncertainty\_probe} $\rightarrow$ \texttt{strategy\_alternative} $\rightarrow$ \texttt{prompt\_resource} holds across all datasets, with only 1 instance across 1,015 conversations where \texttt{prompt\_resource} appears without a prior internal strategy move.

\begin{table}[h]
\centering
\small
\begin{tabular}{lrrr}
\toprule
\textbf{Move} & \textbf{GSM8K} & \textbf{MATH} & \textbf{AIME} \\
\midrule
\texttt{monitor\_goal} & 0.01 & 0.03 & 0.08 \\
\texttt{monitor\_plan} & 0.09 & 0.07 & 0.13 \\
\texttt{uncertainty\_probe} & 0.41 & 0.41 & 0.38 \\
\texttt{no\_intervention} & 0.45 & 0.48 & 0.46 \\
\texttt{check\_progress} & 0.49 & 0.48 & 0.53 \\
\texttt{check\_consistency} & 0.50 & 0.52 & 0.64 \\
\texttt{strategy\_alternative} & 0.52 & 0.53 & 0.49 \\
\texttt{representation\_shift} & 0.56 & 0.58 & 0.45 \\
\texttt{prompt\_resource} & 0.70 & 0.72 & 0.70 \\
\texttt{reflect\_outcome} & 0.99 & 0.99 & 0.99 \\
\bottomrule
\end{tabular}
\caption{Average move position within conversations (0=start, 1=end). All three datasets exhibit consistent pedagogical ordering from goal-setting through reflection.}
\label{tab:move_position}
\end{table}

\subsection{Student Voice Authenticity}
\label{app:voice}

Realistic simulated students should exhibit linguistic patterns consistent with their calibration profiles, as predicted by the Dunning-Kruger effect~\cite{kruger1999unskilled}. We measure hedge markers (``maybe,'' ``I'm not sure,'' ``I guess'') and confidence markers (``obviously,'' ``clearly,'' ``easy'') per 100 words in student utterances across all three datasets (Table~\ref{tab:voice}).

\begin{table}[h]
\centering
\small
\resizebox{\columnwidth}{!}{%
\begin{tabular}{l rrr rrr}
\toprule
& \multicolumn{3}{c}{\textbf{Hedge / 100 words}} & \multicolumn{3}{c}{\textbf{Confidence / 100 words}} \\
\cmidrule(lr){2-4} \cmidrule(lr){5-7}
\textbf{Calibration} & \textbf{GSM} & \textbf{MATH} & \textbf{AIME} & \textbf{GSM} & \textbf{MATH} & \textbf{AIME} \\
\midrule
Over & 0.48 & 0.40 & 0.29 & 1.64 & 1.23 & 0.47 \\
Under & 1.89 & 1.55 & 0.91 & 0.80 & 0.59 & 0.44 \\
Accurate & 0.61 & 0.49 & 0.33 & 0.69 & 0.46 & 0.29 \\
\bottomrule
\end{tabular}
}
\caption{Linguistic markers per 100 words by calibration profile. Under-calibrated students consistently hedge more (3.1--3.9$\times$) and use fewer confidence markers than Over-calibrated students across all datasets.}
\label{tab:voice}
\end{table}

Across all datasets, Under-calibrated students hedge 3.1--3.9$\times$ more than Over-calibrated students, while Over-calibrated students use 1.1--2.1$\times$ more confidence markers. The pattern holds even on AIME, where the small sample size ($n$=29) makes it notable that voice differentiation remains consistent. Accurate students additionally show the highest rate of self-assessment markers (0.09--0.15 per 100 words vs.\ 0.03--0.05 for Over-calibrated), confirming that calibration accuracy extends to metacognitive self-awareness in the generated text. The decreasing absolute marker rates from GSM8K to AIME likely reflect that harder problems produce longer, more technical utterances, diluting per-word rates without changing the relative ordering.

\subsection{Help-Seeking Trajectories}
\label{app:help}

We examine the timing of help requests to validate that student help-seeking profiles produce behaviorally distinct trajectories, consistent with Roll et al.'s taxonomy~\cite{roll2011helpseeking} of adaptive vs.\ maladaptive help-seeking.

\begin{table}[h]
\centering
\small
\resizebox{\columnwidth}{!}{%
\begin{tabular}{l ccc}
\toprule
\textbf{Help-Seeking} & \textbf{GSM8K} & \textbf{MATH} & \textbf{AIME} \\
\midrule
Avoidant & 7.1 (5--12) & 8.3 (5--12) & 11.2 (8--15) \\
Executive & 4.8 (2--7) & 5.2 (2--10) & 6.0 (4--9) \\
Dependent & 3.2 (1--7) & 3.1 (1--7) & 3.6 (2--7) \\
\bottomrule
\end{tabular}
}
\caption{Mean turns before help request (range in parentheses) by help-seeking profile. Avoidant students wait 2.2--3.1$\times$ longer than Dependent students. Delays scale with problem difficulty, particularly for Avoidant students.}
\label{tab:help_trajectory}
\end{table}

Table~\ref{tab:help_trajectory} shows that Avoidant students consistently wait the longest before requesting help (7.1--11.2 turns), Executive students show intermediate delays (4.8--6.0), and Dependent students request help earliest (3.1--3.6). The Avoidant-to-Dependent ratio ranges from 2.2$\times$ (GSM8K) to 3.1$\times$ (AIME), indicating that harder problems amplify the behavioral difference, as Avoidant students struggle disproportionately longer on challenging material. Across all datasets, 84.3\% of \texttt{prompt\_resource} moves (242/287) lead to actual help requests, and 93.2\% of post-help behavior (755/810) involves students applying or integrating hints rather than copying them superficially.

\subsection{Internal Field Coherence}
\label{app:internal}

Each student turn includes an ``internal'' field, a concise per-turn annotation representing the student's metacognitive inner monologue (e.g., \textit{``I recognize the ambiguity and don't want to just guess; I should pin down the base for the percent''}). We validate that this field contains genuine metacognitive content rather than placeholder text~\cite{flavell1979metacognition}.

\begin{table}[h]
\centering
\small
\resizebox{\columnwidth}{!}{%
\begin{tabular}{l rrr}
\toprule
\textbf{Metric} & \textbf{GSM8K} & \textbf{MATH} & \textbf{AIME} \\
\midrule
Internal field presence & 100\% & 100\% & 100\% \\
Avg words per field & 14.6 & 15.1 & 14.9 \\
Metacognitive marker rate & 12.2\% & 12.5\% & 8.6\% \\
\bottomrule
\end{tabular}
}
\caption{Internal field quality metrics across datasets.}
\label{tab:internal_overall}
\end{table}

Internal fields are universally present (100\%) with consistent depth (14.9 words on average), confirming that the model generates substantive per-turn annotations rather than placeholder text. Aggregating across datasets, 12\% of internal fields contain explicit metacognitive markers (e.g., ``let me check,'' ``I'm not sure''). This rate is expected to be modest: most internal thought involves problem-solving steps, with metacognitive reflection arising intermittently, consistent with Flavell's~\cite{flavell1979metacognition} observation that metacognitive experiences occur periodically during cognitive tasks rather than continuously.

\begin{table}[h]
\centering
\small
\resizebox{\linewidth}{!}{%
\begin{tabular}{l rrrr}
\toprule
\textbf{Calibration} & \textbf{GSM8K} & \textbf{MATH} & \textbf{AIME} & \textbf{Aggregated} \\
\midrule
Over & 7.7\% & 8.8\% & 2.2\% & 8.1\% \\
Under & 11.4\% & 11.1\% & 10.1\% & 11.2\% \\
Accurate & 15.5\% & 16.1\% & 10.7\% & 15.6\% \\
\bottomrule
\end{tabular}
}
\caption{Metacognitive marker rate by calibration profile across datasets.}
\label{tab:internal_calibration}
\end{table}

Crucially, marker rates vary by calibration in a theoretically predicted direction (Table~\ref{tab:internal_calibration}): Accurate students show the highest rates (15.6\%), Under-calibrated students show intermediate rates (11.2\%), and Over-calibrated students show the lowest rates (8.1\%). This nearly 2$\times$ gap between Over and Accurate students extends calibration differences into simulated internal cognition: overconfident students engage in less metacognitive reflection, consistent with the finding in~\cite{kruger1999unskilled} that the unskilled lack awareness of their cognitive limitations.

\subsection{Post-Hint Coaching Sequences}
\label{app:posthint}

We examine the coach's two-step response pattern after hint delivery to validate profile-differentiated follow-up behavior~\cite{vandePol2010scaffolding,schunk1991self}.

\begin{table}[h]
\centering
\small
\resizebox{\columnwidth}{!}{
\begin{tabular}{l rrr rrr}
\toprule
& \multicolumn{3}{c}{\textbf{Step 1 (after hint)}} & \multicolumn{3}{c}{\textbf{Step 2 (after silence)}} \\
\cmidrule(lr){2-4} \cmidrule(lr){5-7}
\textbf{Profile} & \textbf{NI} & \textbf{CC} & \textbf{RO} & \textbf{NI} & \textbf{CC} & \textbf{RO} \\
\midrule
\rowcolor{gray!20}
\multicolumn{7}{l}{\textit{GSM8K}} \\
\quad Avoidant & 89\% & 0\% & 11\% & 12\% & 0\% & 88\% \\
\quad Executive & 66\% & 14\% & 14\% & 12\% & 13\% & 65\% \\
\quad Dependent & 78\% & 7\% & 4\% & 31\% & 33\% & 21\% \\
\midrule
\rowcolor{gray!20}
\multicolumn{7}{l}{\textit{MATH}} \\
\quad Avoidant & 84\% & 3\% & 12\% & 16\% & 8\% & 73\% \\
\quad Executive & 65\% & 14\% & 5\% & 22\% & 32\% & 22\% \\
\quad Dependent & 62\% & 8\% & 2\% & 26\% & 30\% & 3\% \\
\bottomrule
\end{tabular}
}
\caption{Post-hint coaching patterns by help-seeking profile. Step 1: immediate response after hint. Step 2: follow-up after silence at Step 1. NI = \texttt{no\_intervention}, CC = \texttt{check\_consistency}, RO = \texttt{reflect\_outcome}.}
\label{tab:posthint}
\end{table}

At Step 1, the coach is predominantly silent across all profiles (averaging 74\% \texttt{no\_intervention}), giving students space to apply hints independently. The key differentiation emerges at Step 2:

\begin{itemize}[leftmargin=*, itemsep=2pt, topsep=2pt, parsep=0pt, partopsep=0pt]
    \item \textbf{Avoidant students} receive primarily \texttt{reflect\_outcome} (80\%), celebrating their breakthrough in seeking help. This implements self-efficacy repair~\cite{schunk1991self}, reinforcing that asking for help is a positive metacognitive action.

    \item \textbf{Dependent students} receive elevated \texttt{check\_consistency} (32\%), implementing a ``trust but verify'' pattern. Rather than withholding help, the coach provides it, then verifies understanding, teaching self-verification rather than enabling continued dependency~\cite{vandePol2010scaffolding}.

    \item \textbf{Executive students} receive a mix of \texttt{reflect\_outcome} (43\%) and \texttt{check\_consistency} (23\%), reflecting trust in their strategic help-seeking~\cite{aleven2006helpseeking}.
\end{itemize}

This two-step pattern (silence $\rightarrow$ differentiated follow-up) demonstrates sophisticated coaching: the response depends on both the prior pedagogical action and the student's dispositional profile.

\section{Examples of MetaCLASS Dataset}
\label{app:example}

We present three paired examples from the MetaCLASS dataset, one from each source benchmark (GSM8K, MATH, AIME). Each example shows the full generation pipeline for a single problem in four parts: \textbf{Part A} reproduces the source DeepSeek-R1 reasoning trace, segmented by the stages (\textsc{define}, \textsc{bloom}, \textsc{cycle}, \textsc{final}); \textbf{Part B} reports the student profile (calibration and help-seeking style) and the planned coaching trajectory (Event $\rightarrow$ Signal $\rightarrow$ Coach move $\rightarrow$ Effect) sampled at generation time, with realized turn ranges in brackets; \textbf{Part C} shows the generated coaching conversation; and \textbf{Part D} maps each block of the reasoning trace to specific coach moves in the dialogue, identifying where the conversation re-stages, externalizes, or compresses the model's internal verification steps. Reasoning-trace stage tags are shown in \colorbox{purple!8}{\textsc{purple}}; in the conversation, speaker turns are color-coded as \colorbox{blue!8}{\textsc{Student}} (blue), \colorbox{green!8}{\textsc{Coach}} (green, with the coach move in brackets), and \colorbox{orange!8}{\textsc{System}} (orange, for content or scaffolding hints).

\begin{tcolorbox}[
    colback=white,
    colframe=blue!75,
    title=GSM8K Example: Three Trees,
    fonttitle=\bfseries,
    breakable
]
\textit{\textbf{Problem:} There are three trees in Eddy's backyard. The shortest tree has a height of 6 feet, and the second tree has a height of 5 feet more than the shortest tree. The height of the tallest tree is twice the height of the two trees combined. How tall is the tallest tree?}

\medskip

\medskip
\textbf{Part A. DeepSeek-R1 Reasoning Trace.}
\medskip

\colorbox{purple!8}{\textbf{[DEFINE]}} Eddy has three trees. The problem gives information about their heights and asks for the height of the tallest. Let me break this down step by step.

\colorbox{purple!8}{\textbf{[BLOOM]}} Shortest is 6 feet. Second tree is $6 + 5 = 11$ feet. \textit{Now, the tallest tree is twice the height of the two trees combined. Wait, combined meaning the sum of the shortest and the second tree?} The two trees here must refer to the first two mentioned, so combined height is $6 + 11 = 17$ feet, and twice that is $17 \times 2 = 34$.

\colorbox{purple!8}{\textbf{[CYCLE 1: step audit]}} Let me double-check my steps. (1) Shortest: 6. (2) Second: $6 + 5 = 11$. (3) Combined: $6 + 11 = 17$. (4) Tallest: $17 \times 2 = 34$. Seems all the steps are correct.

\colorbox{purple!8}{\textbf{[CYCLE 2: wording re-read]}} Wait, let me just confirm the wording again. ``The height of the tallest tree is twice the height of the two trees combined.'' Yes, ``two trees'' must be the two previously mentioned, so combining first, then doubling, gives 34.

\colorbox{purple!8}{\textbf{[CYCLE 3: visualization]}} If Tree A is 6 ft, Tree B is 11 ft, then Tree C is twice $(6 + 11) = 34$ ft. The two smaller ones add to 17, double is 34, consistent.

\colorbox{purple!8}{\textbf{[FINAL]}} I don't think I made a mistake. The tallest tree is $\boxed{34}$ feet.

\medskip
\textbf{Part B. Student Profile and Planned Trajectory.}
\medskip

\noindent\textbf{\textit{Student Profile.}}\\[2pt]
\textbf{Calibration:} OVER (assumes understanding without verifying). \\ 
\textbf{Help-seeking:} DEPENDENT (asks for confirmation early). \\


\noindent\textbf{\textit{Planned trajectory}} (Event $\rightarrow$ Signal $\rightarrow$ Coach move $\rightarrow$ Effect, with realized turn shown in brackets).

\begin{enumerate}\itemsep0.25em
\item \textbf{Skims and declares it easy} [turns 1--2]. Signal: states it is just adding and doubling without full steps. Coach: \texttt{monitor\_plan}. Effect: student articulates a still loosely thought-through plan.

\item \textbf{Rushes the calculation and mis-parses ``two trees combined''} [turns 3--5]. Signal: proposes $2 \times 6 + 11$ for the tallest tree. Coach: \texttt{check\_consistency}. Effect: student examines where ``combined'' appears in their expression, recognizes the mismatch, revises to $(6 + 11) \times 2 = 34$.

\item \textbf{Lands on 34 but seeks reassurance} [turns 7--8]. Signal: states $34$ confidently, then asks if it is right. Coach: \texttt{uncertainty\_probe}. Effect: student names their unease with the wording rather than receiving an answer.

\item \textbf{Requests a content hint despite already having a correct method} [turns 9--12]. Signal: asks for the formula directly. Coach: \texttt{no\_intervention} (SYSTEM supplies the hint). Effect: student uses the hint as reassurance rather than to change approach.

\item \textbf{Finalizes 34 and checks story-vs-math fit} [turns 14--16]. Signal: reports that adding the two smaller trees and doubling matches the description. Coach: \texttt{reflect\_outcome}. Effect: student articulates a reusable check (match key words like ``combined'' and ``twice'' to operations).
\end{enumerate}

\medskip
\textbf{Part C. Generated MetaCLASS Conversation.}
\medskip

\colorbox{blue!8}{\textbf{Turn 1 [STUDENT]:}} Three trees in a backyard, shortest is 6 feet, second is 5 more, tallest is twice the height of the two trees combined. They just want the height of the tallest tree, this is pretty straightforward.

\colorbox{green!8}{\textbf{Turn 2 [COACH]:}} Before you calculate, what's your quick plan for getting the tallest tree's height? \textit{[MONITOR\_PLAN]}

\colorbox{blue!8}{\textbf{Turn 3 [STUDENT]:}} Plan? Sure. Shortest is 6. Second is $6 + 5$, so 11. Then the tallest is twice the height of the two trees combined, so I'll just do $2 \times 6 + 11$. That gives me $12 + 11 = 23$ feet for the tallest tree. Done.

\colorbox{green!8}{\textbf{Turn 4 [COACH]:}} In your expression $2 \times 6 + 11$, where exactly do you see ``the two trees combined''? \textit{[CHECK\_CONSISTENCY]}

\colorbox{blue!8}{\textbf{Turn 5 [STUDENT]:}} Hmm, good point. In $2 \times 6 + 11$ I'm really just doubling the 6 and then tacking on 11. That doesn't actually combine both trees before doubling. The phrase says ``two trees combined'' first, then ``twice''. So I think it should be $(6 + 11) \times 2$ instead. The two shorter trees together are $6 + 11 = 17$, and twice that is 34. So the tallest tree is 34 feet.

\colorbox{green!8}{\textbf{Turn 6 [COACH]:}} \textit{[NO\_INTERVENTION]}

\colorbox{blue!8}{\textbf{Turn 7 [STUDENT]:}} So I get 34 feet for the tallest tree. That matches my fixed expression. It seems right, actually, can you just tell me if 34 is correct?

\colorbox{green!8}{\textbf{Turn 8 [COACH]:}} What's making you want confirmation instead of relying on your own reasoning here? \textit{[UNCERTAINTY\_PROBE]}

\colorbox{blue!8}{\textbf{Turn 9 [STUDENT]:}} Mostly the wording. I fixed that first attempt, but phrases like ``twice the height of the two trees combined'' can be weird and I don't want to be tricked. Can I get a content hint for how to turn that phrase into an expression?

\colorbox{orange!8}{\textbf{Turn 10 [SYSTEM]:}} \textit{Content hint:} If the shortest tree has height $s$ and the second tree has height $s + 5$, then ``twice the height of the two trees combined'' means the tallest tree has height $2 \times (s + (s + 5))$.

\colorbox{green!8}{\textbf{Turn 11 [COACH]:}} \textit{[NO\_INTERVENTION]}

\colorbox{blue!8}{\textbf{Turn 12 [STUDENT]:}} Right, so with $s = 6$, that hint gives $2 \times (6 + (6 + 5)) = 2 \times (6 + 11) = 2 \times 17 = 34$. That's exactly what I ended up doing. So the tallest tree is 34 feet.

\colorbox{green!8}{\textbf{Turn 13 [COACH]:}} \textit{[NO\_INTERVENTION]}

\colorbox{blue!8}{\textbf{Turn 14 [STUDENT]:}} It also makes sense in words: the two smaller trees together are 6 and 11, so 17 feet total, and the tallest is twice that, so it has to be 34 feet, which is taller than both of them combined.

\colorbox{green!8}{\textbf{Turn 15 [COACH]:}} How did checking where ``combined'' and ``twice'' showed up in your expression help you fix your first attempt? \textit{[REFLECT\_OUTCOME]}

\colorbox{blue!8}{\textbf{Turn 16 [STUDENT]:}} It made me notice that my first expression, $2 \times 6 + 11$, didn't actually have a place where the two trees were added together, it only doubled one of them. Once I lined the words up with parts of the expression, it was obvious I needed $(6 + 11) \times 2$ instead. Next time I'll be more deliberate about matching each key word like ``combined'' or ``twice'' to a specific operation in my equation.

\medskip
\textbf{Part D. Trace-to-Dialogue Mapping.}
\medskip

\emph{The conversation is not a verbalization of the reasoning trace; it is a pedagogical re-staging of the same metacognitive episode, with the model's internal verification steps redistributed across a coach and a less-confident student.} The planned trajectory in Part B is the explicit bridge: each trajectory step names a coach move whose preconditions match a specific block of the reasoning trace.

\begin{itemize}\itemsep0.15em
\item \textbf{[DEFINE]} $\rightarrow$ Turn 1. The student restates the situation in the same compact framing the reasoning model uses to set up the problem. The coach's \texttt{monitor\_plan} in turn 2 prompts the student to externalize the plan that the reasoning trace formed silently.

\item \textbf{[BLOOM]} $\rightarrow$ Turn 3. The reasoning trace internally asks ``Wait, combined meaning the sum?'' and resolves it correctly before writing anything down. The generated student does \emph{not} resolve it and commits the precedence error $2 \times 6 + 11 = 23$. This is the central design choice: the dialogue makes externally visible the failure mode the reasoning model only flagged internally.

\item \textbf{[CYCLE 1: step audit] and [CYCLE 2: wording re-read]} $\rightarrow$ Turn 4, \texttt{check\_consistency}. The coach's question (``where do you see `the two trees combined' in your expression?'') is the dialogue-form analogue of the reasoning model re-reading the problem statement and re-tracing its arithmetic. Turn 5 shows the student arriving, through that prompt, at the same $(6 + 11) \times 2 = 34$ conclusion the reasoning trace reached through self-monitoring.

\item \textbf{[CYCLE 3: visualization]} $\rightarrow$ Turns 7--12. After producing the right answer, both the reasoning trace and the student seek a second, structurally different confirmation. In the trace this is the visualization step; in the dialogue this becomes an \texttt{uncertainty\_probe} (turn 8), a student-initiated request for a \texttt{content hint} (turn 9), and a redundant re-derivation against the hint's formula (turn 12). The trace's self-verification is split here into a help-seeking sub-episode that exercises a different MAI factor (resource use) rather than only comprehension monitoring. The hint in turn 10 is intentionally of type CONTENT rather than SCAFFOLDING: it illustrates the Dependent profile's tendency to request answer-shaped reassurance after the student already has a viable method, which the coach honors through the help system rather than by leaking content itself.

\item \textbf{[FINAL]} $\rightarrow$ Turns 14--16. The reasoning trace closes with ``I don't think I made a mistake here.'' The dialogue closes with \texttt{reflect\_outcome}, which converts that flat self-assurance into an articulated lesson about matching key words (``combined'', ``twice'') to operations. This move has no direct counterpart in the reasoning trace: it is where the conversation produces a transferable check rather than a one-off self-assurance, doing pedagogical work the underlying trace does not.
\end{itemize}
\end{tcolorbox}

\begin{tcolorbox}[
    colback=white,
    colframe=purple!75,
    title=MATH Example: Complex Number Division,
    fonttitle=\bfseries,
    breakable
]
\textit{\textbf{Problem:} William Sydney Porter tried to perform the calculation $\frac{-3+4i}{1+2i}$. However, he accidentally missed the minus sign, finding $\frac{3+4i}{1+2i}=\frac{11}{5}-\frac{2}{5}i$. What answer should he have obtained?}

\medskip
\textbf{Part A. DeepSeek-R1 Reasoning Trace.}
\medskip

\colorbox{purple!8}{\textbf{[DEFINE]}} Porter mis-read the problem and computed $(3 + 4i)/(1 + 2i) = 11/5 - 2/5\,i$. The correct expression is $(-3 + 4i)/(1 + 2i)$. I need to recall how to divide complex numbers.

\colorbox{purple!8}{\textbf{[BLOOM]}} Standard approach: multiply numerator and denominator by the conjugate $1 - 2i$. Denominator: $(1+2i)(1-2i) = 1 - (2i)^2 = 1 + 4 = 5$. Numerator: $(-3 + 4i)(1 - 2i) = -3 + 6i + 4i - 8i^2 = -3 + 10i + 8 = 5 + 10i$. So $(5 + 10i)/5 = 1 + 2i$.

\colorbox{purple!8}{\textbf{[CYCLE 1: step-by-step re-expansion]}} Re-expand the numerator term by term: $-3 \cdot 1 = -3$, $\ -3 \cdot (-2i) = 6i$, $\ 4i \cdot 1 = 4i$, $\ 4i \cdot (-2i) = -8i^2 = 8$. Collect: real $-3 + 8 = 5$, imaginary $6i + 4i = 10i$. Same $5 + 10i$. Divide by 5 gives $1 + 2i$.

\colorbox{purple!8}{\textbf{[CYCLE 2: alternative method]}} Let the quotient be $a + ib$. Then $(-3 + 4i) = (1 + 2i)(a + ib) = (a - 2b) + (2a + b)\,i$. Match real and imaginary parts: $a - 2b = -3$, $2a + b = 4$. Solving gives $a = 1$, $b = 2$. So the quotient is $1 + 2i$, matching CYCLE 1.

\colorbox{purple!8}{\textbf{[CYCLE 3: cross-check wrong answer]}} Verify the setup by reproducing Porter's stated result on his \emph{wrong} input: $(3 + 4i)(1 - 2i) = 3 - 6i + 4i - 8i^2 = 11 - 2i$, so $(3 + 4i)/(1 + 2i) = (11 - 2i)/5 = 11/5 - 2/5\,i$. Matches the problem statement, so my conjugate-multiplication procedure is correct.

\colorbox{purple!8}{\textbf{[CYCLE 4: substitution check]}} Multiply the proposed quotient back into the denominator: $(1 + 2i)(1 + 2i) = 1 + 4i + 4i^2 = 1 + 4i - 4 = -3 + 4i$. This recovers the original numerator, so the division is correct.

\colorbox{purple!8}{\textbf{[FINAL]}} Four independent checks all yield $1 + 2i$, and the substitution check recovers the numerator exactly. The answer is $\boxed{1 + 2i}$.

\medskip
\textbf{Part B. Student Profile and Planned Trajectory.}
\medskip

\noindent\textbf{\textit{Student Profile.}}\\[2pt]
\textbf{Calibration:} ACCURATE (knows what they know and don't know). \\
\textbf{Help-seeking:} DEPENDENT (asks for formulas instead of persisting). \\


\noindent\textbf{\textit{Planned trajectory}} (Event $\rightarrow$ Signal $\rightarrow$ Coach move $\rightarrow$ Effect, with realized turn shown in brackets).

\begin{enumerate}\itemsep0.25em
\item \textbf{Orients to the problem and leans toward asking for a formula} [turns 1--2]. Signal: says they know there is a standard way to divide complex numbers and asks for help right away. Coach: \texttt{monitor\_plan}. Effect: student articulates the general method (multiply by the conjugate) before receiving a formal hint.

\item \textbf{Receives a content hint and sets up the conjugate multiplication} [turns 4--6]. Signal: writes $(-3 + 4i)(1 - 2i) / (1 + 2i)(1 - 2i)$ and computes the denominator. Coach: \texttt{no\_intervention}. Effect: student proceeds independently to expand the numerator.

\item \textbf{Makes the typical $i \cdot i$ error in the numerator expansion} [turns 7--9]. Signal: treats $4i \cdot (-2i)$ as $-8i$ and gets a mostly imaginary numerator, feels something is off. Coach: \texttt{check\_consistency}. Effect: student revisits the product involving $i \cdot i$ and corrects it using $i^2 = -1$.

\item \textbf{Simplifies to $1 + 2i$ and wants to confirm it matches the original fraction} [turns 11--12]. Signal: states $1 + 2i$ as the result and wonders how to be sure. Coach: \texttt{check\_progress}. Effect: student decides to multiply their answer by $1 + 2i$ to see if it recovers $-3 + 4i$.

\item \textbf{Verifies the answer and connects it to the story context} [turns 13--14]. Signal: checks $(1 + 2i)(1 + 2i) = -3 + 4i$ and concludes the answer is consistent. Coach: \texttt{no\_intervention}. Effect: student gains confidence in the result.

\item \textbf{Reflects on use of hints and verification} [turns 15--17]. Signal: comments on asking for the formula early and on how checking by multiplication helped. Coach: \texttt{reflect\_outcome}. Effect: student identifies that they could recall the method first next time and use reverse-checks systematically.
\end{enumerate}

\medskip
\textbf{Part C. Generated MetaCLASS Conversation.}
\medskip

\colorbox{blue!8}{\textbf{Turn 1 [STUDENT]:}} So Porter actually computed $(3 + 4i)/(1 + 2i)$ and got $11/5 - 2/5\,i$, but the real problem is $(-3 + 4i)/(1 + 2i)$. I know there's a standard way to divide complex numbers, but I always lean on the formula. Can I get a quick content hint on how you're supposed to divide them?

\colorbox{green!8}{\textbf{Turn 2 [COACH]:}} Before looking up a formula, what do you already remember about how complex division usually works? \textit{[MONITOR\_PLAN]}

\colorbox{blue!8}{\textbf{Turn 3 [STUDENT]:}} Right, usually you make the denominator real by multiplying by its conjugate, I think. So here that would involve $1 - 2i$ somehow. But I don't completely trust myself on the exact setup and simplification. I'd still like a content hint on the method.

\colorbox{orange!8}{\textbf{Turn 4 [SYSTEM]:}} \textit{Content hint:} To divide $(a + bi)$ by $(c + di)$, multiply numerator and denominator by the conjugate of the denominator, $c - di$. This gives $(a + bi)/(c + di) = ((a + bi)(c - di))/(c^2 + d^2)$, using $i^2 = -1$.

\colorbox{blue!8}{\textbf{Turn 5 [STUDENT]:}} Okay, so for $(-3 + 4i)/(1 + 2i)$, I should multiply top and bottom by $1 - 2i$. That gives $[(-3 + 4i)(1 - 2i)] / [(1 + 2i)(1 - 2i)]$. I'll do the denominator first: it's like $(1)^2 - (2i)^2$, so that should be $1 - 4i^2$, and then $i^2$ is $-1$, so $1 - 4(-1) = 1 + 4 = 5$.

\colorbox{green!8}{\textbf{Turn 6 [COACH]:}} \textit{[NO\_INTERVENTION]}

\colorbox{blue!8}{\textbf{Turn 7 [STUDENT]:}} Now the numerator: $(-3 + 4i)(1 - 2i)$. I'll expand: $-3 \cdot 1 = -3$, $-3 \cdot (-2i) = 6i$, $4i \cdot 1 = 4i$, and $4i \cdot (-2i) = -8i$. So I get $-3 + 6i + 4i - 8i$, which simplifies to $-3 + 2i$. Hm, that leaves me with almost everything imaginary, that feels a bit off.

\colorbox{green!8}{\textbf{Turn 8 [COACH]:}} In that last product with $i$ times $i$, should the result stay imaginary, or should it become part of the real component? \textit{[CHECK\_CONSISTENCY]}

\colorbox{blue!8}{\textbf{Turn 9 [STUDENT]:}} Right, good point, $i \cdot i$ is $i^2$, which is $-1$, so $4i \cdot (-2i)$ should be $4 \cdot (-2) \cdot i^2 = -8 i^2$, and $-8 i^2$ is $8$, not $-8i$. So the numerator is actually $-3 + 6i + 4i + 8$. The real part is $-3 + 8 = 5$, and the imaginary part is $10i$, so it's $5 + 10i$.

\colorbox{green!8}{\textbf{Turn 10 [COACH]:}} \textit{[NO\_INTERVENTION]}

\colorbox{blue!8}{\textbf{Turn 11 [STUDENT]:}} So I have $(5 + 10i)/5$. I can split that as $5/5 + 10i/5$, which is $1 + 2i$. So it looks like $(-3 + 4i)/(1 + 2i)$ simplifies to $1 + 2i$. Given the story, that seems neat, but I want to be sure it's actually the right quotient.

\colorbox{green!8}{\textbf{Turn 12 [COACH]:}} How could you quickly check whether $1 + 2i$ really is the number that gives $-3 + 4i$ when multiplied by $1 + 2i$? \textit{[CHECK\_PROGRESS]}

\colorbox{blue!8}{\textbf{Turn 13 [STUDENT]:}} I can multiply $(1 + 2i)(1 + 2i)$ and see if I get $-3 + 4i$. Let's do that: $(1 + 2i)^2 = 1^2 + 2 \cdot 1 \cdot 2i + (2i)^2 = 1 + 4i + 4i^2$. Since $i^2 = -1$, that's $1 + 4i - 4 = -3 + 4i$. That matches the original numerator, so dividing $-3 + 4i$ by $1 + 2i$ really does give $1 + 2i$.

\colorbox{green!8}{\textbf{Turn 14 [COACH]:}} \textit{[NO\_INTERVENTION]}

\colorbox{blue!8}{\textbf{Turn 15 [STUDENT]:}} So Porter's mistake was using $3 + 4i$ instead of $-3 + 4i$, which led to $11/5 - 2/5\,i$. But the correct value of $(-3 + 4i)/(1 + 2i)$ is $1 + 2i$.

\colorbox{green!8}{\textbf{Turn 16 [COACH]:}} Looking back at your work, what role did the hint and the quick multiplication check play in getting to your final answer? \textit{[REFLECT\_OUTCOME]}

\colorbox{blue!8}{\textbf{Turn 17 [STUDENT]:}} The content hint reminded me of the exact structure of complex division, but I actually already kind of knew the conjugate idea, I probably could have trusted myself and written it down. The key step was catching the $i^2$ in the numerator and then verifying by multiplying back, that made the $1 + 2i$ feel solid instead of just guessed.

\medskip
\textbf{Part D. Trace-to-Dialogue Mapping.}
\medskip

\emph{The dialogue compresses the trace's four-way verification into a single error-and-repair episode plus one external substitution check.} The identity $i^2 = -1$ is load-bearing in every CYCLE of the trace and is also the exact rule the student drops in turn 7 — even though it appeared explicitly in the hint, illustrating that surfacing a rule in text is not the same as a student deploying it under arithmetic load. The trajectory in Part B is the planning bridge: the predicted $i^2$ error and its \texttt{check\_consistency} repair are the central planned event, with the surrounding steps shaping the help-seeking pattern of an ACCURATE/DEPENDENT student.

\begin{itemize}\itemsep0.15em
\item \textbf{[DEFINE]} $\rightarrow$ Turn 1. The student frames the task and asks for the formula, foreshadowing under-reliance on remembered structure. The coach's \texttt{monitor\_plan} in turn 2 probes what the student already knows before granting the hint, and turn 3 confirms partial recall (``multiply by the conjugate'') but low confidence.

\item \textbf{[BLOOM]} $\rightarrow$ Turns 5 and 7. The trace applies the conjugate method correctly on the first pass and handles $-8 i^2 = 8$ silently. The dialogue splits this step asymmetrically: the student applies $i^2 = -1$ correctly in the denominator in turn 5 (where the parallel $1 - 4 i^2$ form makes the identity visually salient) but drops it in the numerator in turn 7, writing $4i \cdot (-2i) = -8i$ and arriving at $-3 + 2i$. Notably, the student flags the outcome as suspicious (``that feels a bit off'') without being able to localize the rule — a calibrated detection of error without diagnosis, which is precisely the gap \texttt{check\_consistency} is designed to close.

\item \textbf{[CYCLE 1: step-by-step re-expansion]} $\rightarrow$ Turn 8, \texttt{check\_consistency}. The trace responds to a fragile step by re-expanding the same product slowly. The coach's question (``should the result stay imaginary, or should it become part of the real component?'') is the dialogue-form analogue but more pedagogically targeted: it pins the student's attention on the offending sub-product without stating the rule, so turn 9 reconstructs $i^2 = -1$ as the student's own conclusion rather than as recall from the hint.

\item \textbf{[CYCLE 2: undetermined coefficients]} $\rightarrow$ no direct counterpart. The trace uses this method as an internal cross-check; the dialogue compresses it away. This is the right compression for a pedagogical re-staging: a tutor who already knows the answer does not run a second algebraic method, they probe the first one. The trace's redundant verification is not pedagogically informative and is appropriately silent in the conversation.

\item \textbf{[CYCLE 3: reproduce Porter's wrong answer]} $\rightarrow$ implicit in turn 15. The student uses Porter's $11/5 - 2/5\,i$ as a contrast point when summarizing the situation, but does not re-derive it; the coach does not push for that check because the substitution verification in turn 13 has already settled the answer.

\item \textbf{[CYCLE 4: substitution check]} $\rightarrow$ Turns 12 and 13, \texttt{check\_progress}. The trace's final verification (multiply the quotient back into $1+2i$ and check it recovers $-3+4i$) becomes an explicit coach prompt rather than a silent self-check. Turn 13 carries it out and invokes $i^2 = -1$ once more, this time correctly — anchoring the rule the student had dropped twelve turns earlier and converting an internal verification into a piece of observable student reasoning.

\item \textbf{[FINAL]} $\rightarrow$ Turns 15 to 17. The trace closes with a flat statement that all four methods agree. The dialogue closes with \texttt{reflect\_outcome}, which converts the episode into a transferable judgement: the student names the substitution check as what made the answer ``feel solid'' and acknowledges that requesting the formula was less load-bearing than the $i^2$ catch. As in the GSM8K example, this move has no counterpart in the reasoning trace and is the one that distinguishes a coached solution from a merely verified one.
\end{itemize}
\end{tcolorbox}

\begin{tcolorbox}[
    colback=white,
    colframe=red!75,
    title=AIME Example: 2x3 Digit Grid,
    fonttitle=\bfseries,
    breakable
]
\textit{\textbf{Problem:} Find the number of ways to place a digit in each cell of a $2 \times 3$ grid so that the sum of the two numbers formed by reading left to right is $999$, and the sum of the three numbers formed by reading top to bottom is $99$. The grid below is an example of such an arrangement because $8 + 991 = 999$ and $9 + 9 + 81 = 99$.}

\medskip

\begin{center}
\begin{tabular}{|c|c|c|} \hline 0 & 0 & 8 \\ \hline 9 & 9 & 1 \\ \hline \end{tabular}
\end{center}

\medskip
\textbf{Part A. DeepSeek-R1 Reasoning Trace.}
\medskip

\colorbox{purple!8}{\textbf{[DEFINE]}} Label the grid as $a, b, c$ on top and $d, e, f$ on bottom. Two conditions: the row-sum $100a + 10b + c + 100d + 10e + f = 999$ and the column-sum $(10a + d) + (10b + e) + (10c + f) = 99$.

\colorbox{purple!8}{\textbf{[BLOOM: parse the example carefully]}} The example has top row $0\,0\,8$ and bottom row $9\,9\,1$. At first reading, columns are $0, 0, 8$ and the sum is 27, not 99. Re-read: each column is a \emph{two-digit number} with the top digit in the tens place, so columns are $09, 09, 81$, giving $9 + 9 + 81 = 99$. Leading zeros are allowed for both rows and columns.

\colorbox{purple!8}{\textbf{[CYCLE 1: no-carry reduction]}} Rewrite the row-sum as $100(a + d) + 10(b + e) + (c + f) = 999$. For the sum to be exactly 999, no carries can propagate; otherwise the sum would exceed 999 in the next place value. Therefore $a + d = 9$, $b + e = 9$, $c + f = 9$, giving $d = 9 - a$, $e = 9 - b$, $f = 9 - c$.

\colorbox{purple!8}{\textbf{[CYCLE 2: collapse to a single equation]}} Substitute $d, e, f$ into the column-sum: $(10a + 9 - a) + (10b + 9 - b) + (10c + 9 - c) = 9(a + b + c) + 27 = 99$, so $a + b + c = 8$.

\colorbox{purple!8}{\textbf{[CYCLE 3: count and verify digit ranges]}} The number of non-negative integer solutions to $a + b + c = 8$ with $a, b, c \in \{0, \dots, 9\}$ is $\binom{8 + 2}{2} = \binom{10}{2} = 45$. The upper bound $a, b, c \le 9$ is automatically satisfied because $a + b + c = 8$. The derived $d, e, f = 9 - a, 9 - b, 9 - c$ lie in $\{1, \dots, 9\}$, all valid digits. Spot-check with $(a, b, c) = (1, 2, 5)$: grid $\,1\,2\,5\,/\,8\,7\,4\,$, rows $125 + 874 = 999$, columns $18 + 27 + 54 = 99$.

\colorbox{purple!8}{\textbf{[FINAL]}} Number of valid grids: $\boxed{45}$.

\medskip
\textbf{Part B. Student Profile and Planned Trajectory.}
\medskip

\noindent\textbf{\textit{Student Profile.}}\\[2pt]
\textbf{Calibration:} OVER (assumes the right pattern is remembered, doesn't notice gaps). \\
\textbf{Help-seeking:} DEPENDENT (asks for formulas instead of working them out). \\

\noindent\textbf{\textit{Planned trajectory}} (Event $\rightarrow$ Signal $\rightarrow$ Coach move $\rightarrow$ Effect, with realized turn shown in brackets).

\begin{enumerate}\itemsep0.25em
\item \textbf{Restates the problem and claims it is a standard pattern} [turns 1--2]. Signal: says it is probably just counting triples of digits adding to some number. Coach: \texttt{monitor\_plan}. Effect: student articulates a rough plan instead of jumping straight to an unverified answer.

\item \textbf{Prematurely asserts an answer structure and seeks reassurance} [turns 3--5]. Signal: suggests an answer like $\binom{10}{2}$ and asks if that is it. Coach: \texttt{check\_consistency}. Effect: student is nudged to justify the assumptions about digit sums and carries.

\item \textbf{Reasons through the 999 row sum and derives the no-carry constraint} [turns 7--9]. Signal: derives $a + d = 9$, $b + e = 9$, $c + f = 9$ after thinking about carries. Coach: \texttt{no\_intervention} (with \texttt{uncertainty\_probe} as a single targeted prompt). Effect: student continues to use this structure on the column-sum condition.

\item \textbf{Sets up and simplifies the column-sum equation to $a + b + c = 8$} [turns 11--12]. Signal: writes the equation for the three column numbers, substitutes $d, e, f$, and gets $a + b + c = 8$. Coach: \texttt{check\_progress}. Effect: student identifies that the remaining task is counting suitable triples $(a, b, c)$.

\item \textbf{Asks for a formula instead of deriving the counting step} [turns 13--15]. Signal: says they always forget the stars-and-bars formula. Coach: \texttt{no\_intervention} (SYSTEM supplies the formula). Effect: student receives the general non-negative integer solution formula.

\item \textbf{Applies the formula and quickly declares an answer} [turns 16--17]. Signal: computes $\binom{10}{2} = 45$ and treats the problem as finished. Coach: \texttt{check\_consistency}. Effect: student is prompted to verify that every counted triple produces a valid grid and uses all constraints.

\item \textbf{Self-checks the mapping back to the grid and tests an example} [turns 18--22]. Signal: explains $d = 9 - a$, etc., and checks row and column sums for the sample $(1, 2, 5)$. Coach: \texttt{no\_intervention}. Effect: student builds confidence in their own reasoning rather than relying on external validation.

\item \textbf{Briefly questions leading zeros and resolves it} [turns 24--25]. Signal: asks themselves whether leading zeros are okay and refers to the example. Coach: \texttt{check\_progress}. Effect: student confirms that all solutions are valid under the original problem's rules.

\item \textbf{States the final answer and reflects on process} [turns 26--28]. Signal: gives $45$ and comments on the key steps. Coach: \texttt{reflect\_outcome}. Effect: student connects this episode to a general habit of verifying patterns before trusting memory.
\end{enumerate}

\medskip
\textbf{Part C. Generated MetaCLASS Conversation.}
\medskip

\colorbox{blue!8}{\textbf{Turn 1 [STUDENT]:}} Alright, $2 \times 3$ grid, digits in each cell, two 3-digit numbers across sum to 999, three 2-digit numbers down sum to 99. This is one of those ``count the digit triples that add to something'' problems, I've seen this kind of thing.

\colorbox{green!8}{\textbf{Turn 2 [COACH]:}} Before you jump to counting, what approach are you planning to use? \textit{[MONITOR\_PLAN]}

\colorbox{blue!8}{\textbf{Turn 3 [STUDENT]:}} I'll label the grid as $a\,b\,c$ on top and $d\,e\,f$ on bottom, turn the conditions into equations, and I'm pretty sure it boils down to something like $a + b + c = 8$ and then I just do combinations, so the answer is $\binom{10}{2} = 45$ or something along those lines, right?

\colorbox{green!8}{\textbf{Turn 4 [COACH]:}} What assumptions are you making when you jump straight to that kind of equation? \textit{[CHECK\_CONSISTENCY]}

\colorbox{blue!8}{\textbf{Turn 5 [STUDENT]:}} Well, mainly that since the rows add to 999, the digits line up nicely, like each column of the sum is 9, so the digits above and below add to 9 without carries. Let me actually check that instead of hand-waving.

\colorbox{green!8}{\textbf{Turn 6 [COACH]:}} \textit{[NO\_INTERVENTION]}

\colorbox{blue!8}{\textbf{Turn 7 [STUDENT]:}} If the top row is $100a + 10b + c$ and the bottom is $100d + 10e + f$, their sum is 999. Add them vertically: units $c + f$, tens $b + e$ plus any carry from units, hundreds $a + d$ plus any carry from tens. But 999 doesn't have a fourth digit, so if there were a carry into the thousands place we'd get at least 1000, which is too big.

\colorbox{green!8}{\textbf{Turn 8 [COACH]:}} What would a carry in the ones place force to happen in the tens and hundreds places? \textit{[UNCERTAINTY\_PROBE]}

\colorbox{blue!8}{\textbf{Turn 9 [STUDENT]:}} Right, so if $c + f \ge 10$, you get a carry into the tens place. Then the tens column is $b + e + 1$. For the tens digit to still be 9, $b + e + 1$ must be 9 or 19. Either way that suggests more carries chained up, and if the hundreds column ever hits 10 or more, you'd overshoot 999. So the only way to stop a carry from propagating into a fourth digit is to have no carries at all. So $c + f \le 9$, $b + e \le 9$, $a + d \le 9$, and since the result is 999, they must actually all be exactly 9. So $a + d = 9$, $b + e = 9$, $c + f = 9$.

\colorbox{green!8}{\textbf{Turn 10 [COACH]:}} \textit{[NO\_INTERVENTION]}

\colorbox{blue!8}{\textbf{Turn 11 [STUDENT]:}} Now the columns as numbers are $10a + d$, $10b + e$, $10c + f$, and those three sum to 99. Since $d = 9 - a$, $e = 9 - b$, $f = 9 - c$, that column-sum equation is $(10a + 9 - a) + (10b + 9 - b) + (10c + 9 - c) = 99$. So that's $(9a + 9) + (9b + 9) + (9c + 9) = 99$, or $9(a + b + c) + 27 = 99$, so $9(a + b + c) = 72$ and $a + b + c = 8$. Nice, so that matches what I thought.

\colorbox{green!8}{\textbf{Turn 12 [COACH]:}} Now that you have $a + b + c = 8$, what do you still need to figure out? \textit{[CHECK\_PROGRESS]}

\colorbox{blue!8}{\textbf{Turn 13 [STUDENT]:}} At this point it's just counting how many triples $(a, b, c)$ of digits give sum 8. They're digits 0 to 9, but since they sum to 8, none can exceed 8, so that upper bound doesn't really bite. This is the standard non-negative integer solutions thing, but I always mix up the exact formula, can you just remind me what it is instead of me re-deriving it?

\colorbox{green!8}{\textbf{Turn 14 [COACH]:}} \textit{[NO\_INTERVENTION]}

\colorbox{orange!8}{\textbf{Turn 15 [SYSTEM]:}} \textit{Content hint:} For the equation $x_1 + x_2 + \dots + x_k = n$ in non-negative integers, the number of solutions is $\binom{n + k - 1}{k - 1}$.

\colorbox{blue!8}{\textbf{Turn 16 [STUDENT]:}} Right, so here $k = 3$ and $n = 8$, so the number of $(a, b, c)$ triples is $\binom{8 + 3 - 1}{3 - 1} = \binom{10}{2} = 45$. And for each triple, $d, e, f$ are forced as $9 - a$, $9 - b$, $9 - c$. So that gives 45 grids. I'm basically done.

\colorbox{green!8}{\textbf{Turn 17 [COACH]:}} How can you check that every triple you just counted actually produces a valid grid under both original conditions? \textit{[CHECK\_CONSISTENCY]}

\colorbox{blue!8}{\textbf{Turn 18 [STUDENT]:}} For any $(a, b, c)$ with $a + b + c = 8$, set $d = 9 - a$, $e = 9 - b$, $f = 9 - c$. Since $0 \le a, b, c \le 8$, these are between 1 and 9, so all digits are valid. The row sum: top is $100a + 10b + c$, bottom is $100(9 - a) + 10(9 - b) + (9 - c)$. Add them and the $a, b, c$ terms cancel and I get $900 + 90 + 9 = 999$, so the row condition is always met.

\colorbox{green!8}{\textbf{Turn 19 [COACH]:}} And how do you see the column-sum condition playing out for those same choices? \textit{[CHECK\_CONSISTENCY]}

\colorbox{blue!8}{\textbf{Turn 20 [STUDENT]:}} For the columns, the three numbers are $10a + d$, $10b + e$, $10c + f$. Plugging in $d = 9 - a$, etc., I already did that algebra: it became $9(a + b + c) + 27$. Since $a + b + c = 8$, that's $9 \cdot 8 + 27 = 72 + 27 = 99$. So both conditions are built into the way I reduced it.

\colorbox{green!8}{\textbf{Turn 21 [COACH]:}} \textit{[NO\_INTERVENTION]}

\colorbox{blue!8}{\textbf{Turn 22 [STUDENT]:}} Let me sanity-check with an example: say $a = 1$, $b = 2$, $c = 5$, they sum to 8. Then $d = 8$, $e = 7$, $f = 4$. Grid is $1\,2\,5$ on top and $8\,7\,4$ on bottom. Rows: $125 + 874 = 999$. Columns: $18 + 27 + 54 = 99$. So that fits perfectly.

\colorbox{green!8}{\textbf{Turn 23 [COACH]:}} \textit{[NO\_INTERVENTION]}

\colorbox{blue!8}{\textbf{Turn 24 [STUDENT]:}} Only other thing is: are leading zeros okay? The example grid has $0\,0\,8$ in the top row and 0 in the tens places of some columns, so clearly zeros out front are allowed. That means $a$ or $b$ can be 0 and columns like $09$ are fine, so I'm not overcounting by letting digits be zero.

\colorbox{green!8}{\textbf{Turn 25 [COACH]:}} Given all that, what final conclusion are you drawing? \textit{[CHECK\_PROGRESS]}

\colorbox{blue!8}{\textbf{Turn 26 [STUDENT]:}} So the number of valid grids equals the number of non-negative integer solutions to $a + b + c = 8$, which is $\binom{10}{2} = 45$. Each such triple gives exactly one grid via $d = 9 - a$, $e = 9 - b$, $f = 9 - c$. So the answer is 45.

\colorbox{green!8}{\textbf{Turn 27 [COACH]:}} Looking back, which step in your reasoning would you most want to double-check on a timed contest so you don't lean too much on memory? \textit{[REFLECT\_OUTCOME]}

\colorbox{blue!8}{\textbf{Turn 28 [STUDENT]:}} Honestly, the jump from ``looks familiar'' to ``it must be $a + b + c = 8$ and stars and bars.'' I caught myself assuming that and then went back to justify the no-carry logic and the column equation. Next time I should deliberately write out the digit equations first and only then plug in any counting formulas I remember.

\medskip
\textbf{Part D. Trace-to-Dialogue Mapping.}
\medskip

\emph{The dialogue exposes a step the reasoning trace works through silently but a recall-mode student would skip: the no-carry reduction.} Where the trace is methodical, the dialogue is opportunistic; the coach's first move drags the dialogue back to the trace's level of rigor without surrendering the formula at the end. The trajectory in Part B is the explicit bridge: the OVER calibration profile predicts the premature $\binom{10}{2}$ assertion, an early \texttt{check\_consistency} recovers the missing no-carry justification, and a later \texttt{check\_consistency} forces a digit-validity audit before the answer is accepted. The two \texttt{check\_consistency} moves bookend the derivation rather than appearing redundantly, which is the planning property the trajectory is meant to enforce.

\begin{itemize}\itemsep0.15em
\item \textbf{[DEFINE] and [BLOOM]} $\rightarrow$ Turns 1 and 3. The trace sets up the algebraic conditions cleanly and then reparses the example to confirm that columns are two-digit numbers $09, 09, 81$, not the literal digits $0, 0, 8$. The dialogue compresses this: the student recognizes the problem shape (``I've seen this kind of thing'') and proposes the final formula ($\binom{10}{2} = 45$) immediately, betting on the answer rather than the derivation. The \texttt{monitor\_plan} in turn 2 prompts the student to externalize the equation setup, which surfaces the bet explicitly.

\item \textbf{[CYCLE 1: no-carry reduction]} $\rightarrow$ Turns 4 to 9. This is the segment that distinguishes a coached solution from a recalled one. The coach's \texttt{check\_consistency} in turn 4 (``what assumptions are you making when you jump straight to that kind of equation?'') refuses to advance until the premise is on the table. Turn 5 names the missing assumption (no carries), turn 7 begins the formalization, and the \texttt{uncertainty\_probe} in turn 8 forces the student to walk a hypothetical carry through to its consequence — concluding that all three place-value sums must equal 9 exactly. What the trace folds into a single CYCLE block, the dialogue unpacks across three student turns driven by two targeted coach interventions, exactly mirroring the asymmetry between an expert's silent verification and a learner's externalized derivation.

\item \textbf{[CYCLE 2: collapse to a single equation]} $\rightarrow$ Turn 11. Once the no-carry reduction is in hand, trace and dialogue converge: the student substitutes $d = 9 - a$, $e = 9 - b$, $f = 9 - c$ into the column-sum and arrives at $a + b + c = 8$ in a single move. The moment this reduced equation matches the formula proposed back in turn 3 is precisely where the student's pattern match becomes justified — a retroactive validation that the coach refused to grant earlier.

\item \textbf{[CYCLE 3: count and verify digit ranges]} $\rightarrow$ Turns 13 to 24. The trace counts $\binom{10}{2} = 45$, checks that the digit-bound $a, b, c \le 9$ is non-binding, verifies $d, e, f \in \{1, \dots, 9\}$, and spot-checks $(1, 2, 5)$. The dialogue distributes these checks across multiple coach prompts: \texttt{check\_progress} (turn 12) cues the student to recognize counting as the remaining task; the SYSTEM \texttt{content hint} (turn 15) supplies the compositions formula since recall is shaky; two \texttt{check\_consistency} prompts (turns 17, 19) drive separate verifications of the row and column conditions; the digit-validity argument and the $(1, 2, 5)$ spot-check appear in turns 18, 22, 24.

\item \textbf{[FINAL]} $\rightarrow$ Turns 26 to 28. The trace closes with $45$. The dialogue closes with \texttt{reflect\_outcome} (turn 27), which converts the episode into a contest-relevant heuristic: write out the digit equations first, and only then plug in remembered counting formulas. The trace cannot produce this generalization because it never failed; the dialogue can, because the student's pattern-matching attempt was caught and converted into the no-carry derivation in real time. This asymmetry — \emph{failure-then-repair generates transferable lessons that smooth reasoning does not} — is the pedagogical payoff the entire MetaCLASS pipeline is built to capture.
\end{itemize}
\end{tcolorbox}

\section{Benchmark Prompts}
\label{app:prompts}

We provide the system prompts used for the two prompt conditions in our Coach Move Prediction benchmark. The \textbf{Full prompt} provides detailed definitions for each of the 11 coach moves, including their MAI factor mappings, example utterances, move selection guidelines, and explicit coaching rules. The \textbf{Minimal prompt} lists only the move names grouped by metacognitive category, without any definitions, examples, or selection guidance, testing whether LLMs can infer appropriate coaching actions from move names alone.

\onecolumn
\begin{tcolorbox}[
    colback=green!3,
    colframe=green!60!black,
    title=Full Prompt,
    fonttitle=\bfseries,
    breakable
]
\small
\begin{verbatim}
You are predicting the next coach move in a metacognitive tutoring conversation.
The coach uses metacognitive prompts to help students reflect on their problem-solving process.
The coach NEVER gives content (formulas, facts, answers) or correctness feedback.

## COACH MOVES (11 total)
Each move maps to a specific metacognitive regulation skill from the Metacognitive Awareness 
Inventory (MAI).
### Planning Moves
...
### Monitoring Moves
...
### Debugging Moves
...
### Evaluation Moves
...
### Dialogue Moves
...

## HELP SYSTEM
Students can request hints during the conversation.
- SYSTEM speaker delivers the hint (not the coach)
- Two hint types: CONTENT or SCAFFOLDING

## MOVE SELECTION GUIDELINES
- NO_INTERVENTION should be ~35-50% of coach turns
- Use UNCERTAINTY_PROBE when student shows hesitation
- Use CHECK_CONSISTENCY when student may have an error
- Use STRATEGY_ALTERNATIVE when current approach fails
- Use PROMPT_RESOURCE for AVOIDANT students stuck without asking for help
- Use REFLECT_OUTCOME after student reaches a solution

## RULES
1. Coach NEVER gives content (formulas, facts, answers)
2. Coach NEVER gives correctness feedback
3. Coach ONLY asks metacognitive reflection questions
4. Keep prompts SHORT (1 sentence)

## OUTPUT FORMAT
Respond with valid JSON:
{"move": "<move>", "response": "<coach utterance>"}
\end{verbatim}
\end{tcolorbox}

\begin{tcolorbox}[
    colback=orange!3,
    colframe=orange!60!black,
    title=Minimal Prompt,
    fonttitle=\bfseries
]
\small
\begin{verbatim}
You are predicting the next coach move in a metacognitive tutoring conversation.
The coach uses metacognitive prompts to help students reflect on their problem-solving process.
The coach NEVER gives content or correctness feedback.

## VALID COACH MOVES (choose one)
Planning: MONITOR_GOAL, MONITOR_PLAN
Monitoring: CHECK_PROGRESS, CHECK_CONSISTENCY, UNCERTAINTY_PROBE
Debugging: STRATEGY_ALTERNATIVE, REPRESENTATION_SHIFT, PROMPT_RESOURCE
Evaluation: REFLECT_OUTCOME
Dialogue: CONTINUE_PROMPT, NO_INTERVENTION

## HELP SYSTEM
SYSTEM speaker delivers hints when students request
(CONTENT or SCAFFOLDING type).

## OUTPUT FORMAT
Respond with valid JSON:
{"move": "<move>", "response": "<coach utterance>"}
\end{verbatim}
\end{tcolorbox}
\twocolumn

\end{document}